\documentclass[twocolumn,showpacs,amsmath,amssymb,superscriptaddress,a4paper]{revtex4}
\usepackage{graphicx}
\usepackage{dcolumn}%
\usepackage{bm}
\usepackage{longtable}
\usepackage{epsfig}
\usepackage{amsmath}
\usepackage{units}
\usepackage{todonotes}
\usepackage{color}
\usepackage{braket}

\newcommand{\mub}{\ensuremath{\mu_B}}
\renewcommand{\vec}[1]{\bm{#1}}

\newcommand{\PH}[2]{$\textnormal{#1}_{\textnormal{#2}}$}
\newcommand{\HBCOx}{HgBa$_2$CuO$_{4+\delta}$}
\newcommand{\HBCO}{HgBa$_2$CuO$_{4}$}

\newcommand{\LSCOx}{La$_{2-x}$Sr$_x$CuO}

\begin{document}
\title{Quasi-static magnetoelectric multipoles as the order parameter for the pseudo-gap phase in cuprate superconductors}
\author{M. Fechner}
\email{michael.fechner@mat.ethz.ch}
\affiliation{Materials Theory, ETH Zurich,  Wolfgang-Pauli-Strasse 27, 8093 Z\"{u}rich, Switzerland }
\author{M. J. A. Fierz}
\affiliation{Materials Theory, ETH Zurich,  Wolfgang-Pauli-Strasse 27, 8093 Z\"{u}rich, Switzerland }
\author{F. Th\"{o}le}
\affiliation{Materials Theory, ETH Zurich,  Wolfgang-Pauli-Strasse 27, 8093 Z\"{u}rich, Switzerland }
\author{U. Staub}
\affiliation{Swiss Light Source, Paul Scherrer Institut, CH-5232 Villigen PSI, Switzerland.}
\author{N. A. Spaldin}
\affiliation{Materials Theory, ETH Zurich,  Wolfgang-Pauli-Strasse 27, 8093 Z\"{u}rich, Switzerland }

\begin{abstract}
We introduce a mechanism in which coupling between fluctuating spin magnetic dipole moments and polar optical phonons leads to a non-zero ferroic ordering of quasi-static magnetoelectric multipoles. Using first-principles calculations within the LSDA$+U$ method of density functional theory, we calculate the magnitude of the effect for the prototypical cuprate superconductor, \HBCO. We show that our proposed mechanism is consistent, to our knowledge, with all experimental data for the onset of the pseudo-gap phase and therefore propose the quasi-static magnetoelectric multipole as a possible pseudo-gap order parameter. Finally, we show that our mechanism embraces some key aspects of previous theoretical models, in particular the description of the pseudo-gap phase in terms of orbital currents. 
\end{abstract}

\maketitle
\section{Introduction}
The partial persistence of the superconducting gap above the superconducting transition temperature in the under-doped high-$T_c$ cuprate superconductors has long been acknowledged to be key in understanding the nature of pairing in the superconducting state \cite{Varma:2010}. Following the first observation of its signature in the spin-lattice relaxation rate in nuclear magnetic resonance (NMR) experiments \cite{Warren_et_al:1989}, it has been unclear for many years whether this so-called {\it pseudo-gap} region forms a distinct phase, with a line of phase transitions into the neighboring strange-metal phase, or whether doping instead causes a continuous increase in the Fermi surface and a corresponding gradual crossover to strange metallicity. Indeed, the usual signatures of a phase transition such as changes in known symmetries or singularities in susceptibilities, remain elusive. The importance of the distinction -- the phase transition scenario naturally supports the proposed existence of a zero-kelvin quantum critical point within the superconducting dome -- has motivated intensive and ongoing research. 

A variety of often technically challenging experiments, including spin-polarized elastic neutron diffraction \cite{Fauque:2006bo}, ultra-sound measurements \cite{Shekhter_et_al:2013}, nuclear magnetic (NMR) and quadrupolar (NQR) resonance \cite{Strassle:2011dr,Tokunaga:1999df} and muon spin resonance ($\mu$SR) \cite{Sonier_et_al:2001} now point to the existence of a phase transition, and the focus of attention has shifted to determining the nature of the ``hidden'' order parameter. The experimental situation regarding the identity of the order parameter is unconverged, with different experiments giving apparently conflicting results, at least when they are interpreted in terms of an ordering that involves magnetic dipole moments. The key findings are as follows:

A clear indication that the pseudo-gap ordering has a magnetic origin comes from spin-polarized elastic neutron diffraction which finds differences between the non-spin-flip and spin-flip channels below $T^*$ in under-doped YBa$_2$Cu$_3$O$_{6+x}$ (YBCO) \cite{Fauque:2006bo}. For over-doped samples outside of the pseudo-gap regime, in contrast, the spin-flip and non-spin-flip signals are both flat indicating zero magnetic intensity at all temperatures. The observed increase in the spin-flip signal when the neutron polarization is parallel to the wavevector of the Q=(0,1,1) Bragg peak indicates that the time-reversal symmetry breaking magnetic order lies on top of the nuclear Bragg peaks and so preserves translational symmetry, thereby excluding likely antiferromagnetic orderings of the Cu ions. No scattering is observed when the in-plane component of the scattering vector is zero, however, indicating, apparently inconsistently, zero net ferromagnetic moment. An interpretation of the scattering in terms of magnetic dipole moments requires that any possible magnetic dipole moments be canted at \unit[45]{$^{\circ}$} to the Cu-O plane, with a magnitude of $\sim$\unit[0.1]{\mub}. Some ARPES measurements also point to time-reversal symmetry breaking, with measurements on Bi$_2$Sr$_2$CaCu$_2$O$_{8-\delta}$ (Bi-2212), for example, yielding different photocurrents for left- and right-circularly polarized photons below the pseudo-gap onset temperature in the under-doped regime \cite{Kaminski:2002uj}. The interpretation of this result as further evidence of time-reversal symmetry breaking remains hotly disputed however (see for example Refs.~\onlinecite{Borisenko_et_al:2004,Arpiainen/Bansil/Lindroos:2009,Norman/Kaminski/Rosenkranz:2010}).

Apparently inconsistent with the absence of ferromagnetism indicated by the neutron data is the observation of a signal below $T^*$ in magneto-optic Kerr effect measurements on YBCO \cite{Xia:2008cc}.  While such a Kerr signal provides additional confirmation of time-reversal symmetry breaking, it is usually indicative of ferromagnetic ordering which the neutron measurements exclude. The signal strength, $\sim$\unit[1]{$\mu$~rad}, is about four orders of magnitude smaller than the response in a typical ferromagnetic transition metal oxide. In addition, it shows an unusual temperature dependence, which implies that the source of the Kerr rotation is not the primary order parameter for the phase transition. Intriguingly, field-training data suggest that time-reversal symmetry is already broken above $T^*$ by an order parameter that interacts with that of the pseudo-gap phase, even though the Kerr rotation is below the detection limit in that temperature range.

Results from $\mu$SR spectroscopy, which is sensitive to local magnetic fields at the site of muon implantation in a sample, are also contradictory. Early $\mu$SR experiments on YBCO \cite{Sonier_et_al:2001}, and subsequent work on \LSCOx\ (LSCO) \cite{MacDougall:2008br} gave upper bounds on internal static magnetic fields of \unit[0.01]{mT} and \unit[0.02]{mT} respectively; these are three orders of magnitude smaller than the fields that should be present if the spin-polarized neutron scattering were caused by a magnetic dipolar ordering. It was pointed out, however, that any order fluctuating with a timescale faster than $10^{-6}$ s would be averaged to zero in a $\mu$SR experiment \cite{MacDougall:2008br}. Later experiments on YBCO \cite{Sonier:2009fp} even suggested an impurity phase origin for the weak magnetic signal. The small internal magnetic fields obtained in the muon measurements are supported by NMR \cite{Strassle:2008ku} and NQR \cite{Strassle:2011dr} studies on under-doped YBCO, which place upper limits of \unit[0.15]{mT} and \unit[0.07]{mT} on the static local magnetic fields at the Y and Ba sites respectively, and of \unit[0.7]{mT} at both sites for rapidly (faster than $10^{-11}$s) fluctuating fields. 

An important theoretical description of the pseudo-gap phase, which is both conceptually appealing and consistent with many of the experimental observations, is the so-called {\it orbital-current} model \cite{Varma:1997}. In this model, oppositely oriented electron-current loops flow identically within each unit cell producing intra-unit cell antiferromagnetically aligned magnetic moments. The current loops introduce magnetic moments that can be arbitrarily small depending  on the magnitudes of the currents, and the ordering breaks time-reversal symmetry while retaining translational invariance. In its existing form, however, the model is seemingly inconsistent with the NMR, NQR and $\mu$SR results \cite{Strassle:2008ku,Sonier:2009fp,Strassle:2011dr}, and a microscopic origin of the behavior is not obvious. 

A recent theoretical analysis demonstrating that neutrons are deflected by magnetoelectric multipoles \cite{Lovesey:2014gp} led to the intriguing suggestion that a ferroic ordering of magnetic quadrupoles on the Cu ions in YBCO could have a symmetry consistent with that inferred from the neutron scattering measurements \cite{Lovesey:2014vm}. The magnetoelectric multipoles, of which the magnetic quadrupoles are one type, form the second-order term in the multipole expansion of the interaction energy of the magnetization density with a magnetic field \cite{Spaldin_et_al:2013} (the magnetic dipole forms the first-order term) and so break time-reversal symmetry without carrying a magnetic dipole moment. Ref.~\onlinecite{Lovesey:2014vm} discounted evidence from neutron scattering, specifically the observed structure factor of the L=0 Bragg peak, which requires canted moments when interpreted in terms of magnetic dipole moments \cite{ManginThro_et_al:2015}. As a result their proposed ordering of $z^2$ quadrupoles, which leads to a $Cm'm'm'$ space group, is inconsistent with experiment. The concept remains relevant, however, for the lower $C2/m'$ symmetry that captures all of the neutron data. Such a ``ferromagnetoelectric'' order parameter is also supported by a symmetry analysis motivated by the Kerr effect results \cite{Orenstein:2011}, which identified magnetic point group symmetries -- $2/m'$, $m'm'm'$, $2m'm'$ -- that could generate polarization rotation without a net ferromagnetism via the magnetoelectric effect. How to achieve such a ferromagnetoelectric order microscopically, however, in the absence of an additional magnetic dipolar order is far from obvious.

Here we use first-principles calculations based on density functional theory to demonstrate the existence and calculate the magnitudes of magnetoelectric multipoles in the prototypical cuprate superconductor \HBCOx. We show that all existing experimental reports of the pseudo-gap phase are consistent, to our knowledge, with its order parameter being the ferroic ordering of such magnetoelectric multipoles. We provide a mechanism for the ordering of the magnetoelectric multipoles in the absence of magnetic dipolar ordering that is mediated by coupling between fluctuating spin dipole moments and optical phonons. We show that our analysis embraces many aspects of, and provides a microscopic justification for, the theoretical models of  Refs.~\onlinecite{Varma:1997} and \onlinecite{Lovesey:2014vm} while also accounting for the experimental observations that they do not capture. Finally, we present proposals for experiments that could directly verify or disprove our proposed mechanism.

\section{Mercury Barium Copper Oxide}

We choose as the subject of our study the model high-$T_c$ cuprate \HBCO\ (Hg-1201) \cite{Barisic_et_al:2008}, which shows the usual cuprate phase diagram, with its characteristic superconducting dome \cite{Yamamoto/Hu/Tajima:2000} as a function of hole doping. It has a simple tetragonal $P4/mmm$ crystal structure (Fig.~\ref{fig2} (a)) which, importantly for our study, contains a single Cu-O plane per unit cell, minimizing the possible magnetic ordering combinations that preserve translational symmetry. In particular, since it contains only one Cu ion per unit cell, any antiferromagnetic ordering of magnetic moments on the Cu ions is excluded. The ideal parent compound, with its +2 oxidation state for Cu, has to our knowledge not been achieved, since there is always a non-zero amount of hole doping through incorporation of interstitial oxygen. The oxygen interstitial site is in the Hg plane, equidistant from the Hg ions, and so doping does not introduce disorder into the Cu-O planes responsible for the superconductivity \cite{Huang:1995tn}. A range of hole concentrations from $\delta = 0.05 - 0.25$ (in units of number of holes per Cu ion) has been reported, with a $T_c$ of 97 K at optimal doping ($\delta=0.15$). 

The existence and doping dependence of the pseudo-gap phase in Hg-1201 is confirmed from thermoelectric power measurements \cite{Yamamoto/Hu/Tajima:2000}, $^{63}$Cu \cite{Itoh_et_al:1996} and $^{17}$O \cite{Mounce_et_al:2013} NMR, photoemission \cite{Uchiyama_et_al:2000} and spin-polarized neutron diffraction \cite{Li_et_al:2008,Li:2012is} to have the same overall behavior as in the other cuprate superconductors. As mentioned above, the neutron data for all cuprates point to an ordering in the pseudo-gap phase that breaks time-reversal symmetry but not the translational symmetry of the lattice; in the case of Hg-1201 with one Cu ion per unit cell this fact, combined with the absence of ferromagnetism, excludes the presence of ordered magnetic dipole moments on the Cu ions. A recent meta-analysis of all available neutron studies on Hg-1201 \cite{Lovesey:2015wb} concluded that the neutron data are consistent with the $Cm'm'm'$ magnetic space group which prohibits ordered magnetic dipole moments at the Cu sites and also breaks space-inversion symmetry. Note that experimental information for the magnetic structure factor of the $L=0$ peak is not yet available, due to the strong scattering from the nuclear $L=0$ peak \cite{BourgesPC}, so it is not possible to know whether the actual magnetic symmetry is lowered to $C2/m'$ (which anyway also breaks space-inversion symmetry and prohibits ordered magnetic dipoles on the Cu sites) as in the YBCO case. Also relevant for our discussion is the finding, using inelastic neutron scattering \cite{Li:2012is} and from interpretation of optical scattering \cite{Yang_et_al:2009}, of the onset at $T^*$ of a weakly dispersive collective spin excitation at around \unit[40]{meV}. We mention finally that charge density wave correlations have been measured using Cu $L_3$-edge resonant X-ray scattering at a lower temperature than  $T^*$ \cite{Tabis:2014kb} and interpreted in terms the build-up of significant dynamic antiferromagnetic correlations at $T^*$.

While the detailed analysis we present here, particularly the numerical density functional theory study, is specific to Hg-1201, we believe that our conclusions are generally applicable to the whole family of high-$T_c$ cuprates.

\section{Magnetoelectric multipoles in {H\lowercase{g}-1201}}

The magnetoelectric multipoles form the second order contributions to the multipole expansion of the energy of a general magnetization density interacting with a general magnetic field \cite{Spaldin_et_al:2013}. (The first order contribution comes from the magnetic dipole moment, and a true magnetic monopole would give a zeroth order contribution). Their operators are formed from a product of a position operator, $\hat{\vec{r}}$, and a magnetization operator, $\hat{\vec{\mu}}$ and therefore they are only non-zero in materials that have both broken time-reversal and space-inversion symmetry. As a result, materials with non-zero magnetoelectric multipoles also exhibit a linear magnetoelectric response in which an electric field induces a proportional magnetization and vice versa. In most magnetoelectrics discussed to date, the source of the magnetoelectric multipolar ordering is an antiferromagnetic ordering of magnetic dipoles which breaks simultaneously both time-reversal and space-inversion symmetry, although inversion symmetry can also be broken separately for example by a polar structural distortion. 

\begin{figure}[tb]
\includegraphics[width=1\columnwidth]{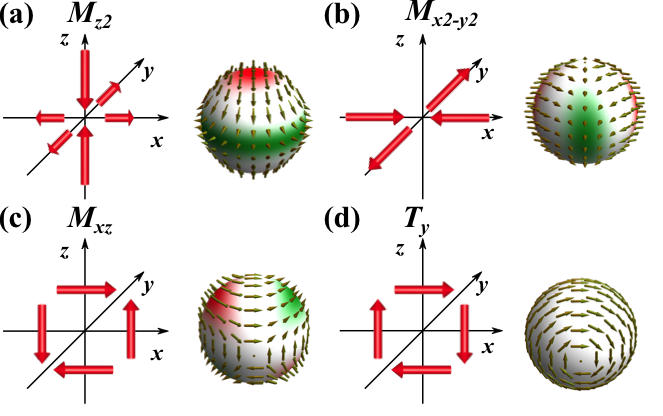}
\caption{\label{fig1} Pattern of local magnetic dipole moments, $\mu$ that generate magnetoelectric multipoles. (a),(b) and (c) show the cases of magnetic quadrupoles with $M_{z^2}$, $M_{x^2 - y^2}$ and $M_{z^2}$ symmetry, respectively. (d) pattern of local magnetic dipole moments that generates a toroidal moment $\vec{T}_y$.}
\end{figure}

The chemical unit cell of Hg-1201 shown in Fig.~\ref{fig2} (a) is clearly centrosymmetric with an inversion center at the Cu site. In the absence of ordered magnetic moments it is also time-reversal symmetric, so all magnetoelectric multipoles at all atoms are zero. Here we explore scenarios in which non-zero magnetoelectric multipoles can emerge from such a high temperature centrosymmetric paramagnetic state and eventually represent a primary order parameter for a symmetry-lowering phase transition. For conciseness, we focus particularly on the magnetic quadrupole which is a second-rank tensor given by \cite{Spaldin_et_al:2013}
\begin{equation}\label{eq:quad}
M_{ij}=\frac{1}{2}\int \left[ r_i \mu_j + r_j \mu_i - \frac{2}{3} \vec{r}\cdot\vec{\mu}(\vec{r}) \delta_{i,j}\right] d^3r \quad, 
\end{equation}
with energy of interaction, $E_{\rm int}$, with a magnetic field $\vec{H}$ determined by the field gradients according to 
\begin{equation}
E_{\rm int} =  - M_{ij} \left(\partial_{i} H_{j} + \partial_{j} H_{i}\right)_{\vec{r} = 0} \quad.
\end{equation}
Such a magnetic quadrupole can be generated by a pattern of local magnetic dipole moments, such as that shown in Fig.~\ref{fig1} (a), which might represent spin moments on atoms within a unit cell. The pattern in Fig.~\ref{fig1} (a) shows a $z^2$ magnetic quadrupole, $M_{z^2}$, whose magnitude is straightforwardly obtained from Eqn.~\eqref{eq:quad} by summing over the atoms and replacing $\vec{\mu}(\vec{r})$ by the local moments $\mu$; the values per unit cell are converted to a macroscopic ``quadrupolization'' by dividing by the unit cell volume \cite{Spaldin_et_al:2013}. Note that the pattern of ordered dipole moments breaks the inversion symmetry. In addition to this ``unit-cell'' contribution to the total quadrupolization formed from the separation of magnetic dipoles at the unit-cell length scale, there is also an ``atomic-site'' contribution to the total quadrupolization, also illustrated in Fig.~\ref{fig1} (a), which arises from the magnetization texture within a sphere centered at individual atoms, provided that their site symmetry is appropriate. Fig.~\ref{fig1} (b) shows the analogous unit-cell and atomic-site contributions to the $x^2 - y^2$ magnetic quadrupole, $M_{x^2-y^2}$, and (c) the $xz$ magnetic quadrupole, $M_{xz}$. In Tab.~\ref{tab:mag_symmetry} we list the point group symmetry on the Cu site caused by the occurrence of these quadrupoles either individually or in various combinations on the Cu ion. We see that both $M_{z^2}$ and $M_{x^2-y^2}$ have $m'm'm'$ symmetry, and any combination of either the $z^2$ or $x^2 - y^2$ quadrupole with either the $xz$ or $yz$ quadrupole yields $2/m'$ symmetry. Symmetry analysis of the toroidal magnetoelectric multipoles shows that a toroidal moment oriented in the $y$ direction, $\vec{T}_y$, on the Cu site has the same $m'mm$ symmetry as $M_{xz}$ and $M_{yz}$, so its combination with the $z^2$ or $x^2 - y^2$ quadrupole also yields the proposed $2/m'$ symmetry. We illustrate such a $\vec{T}_y$ toroidal moment in Fig.~\ref{fig1} (d). (Note that the toroidal moments are sometimes called anapoles in the literature.) The magnetoelectric monopole has the same $m'm'm'$ symmetry as the $z^2$ quadrupole and so is always simultaneously allowed. Since magnetoelectric monopoles have been shown to not interact with neutrons, however, \cite{Lovesey:2014gp} we do not discuss them here.

\subsection{Possibilities for generating magnetoelectric multipoles from magnetic dipolar ordering}
First, we analyze the possibilities within the higher symmetry $Cm'm'm'$ \cite{Lovesey:2014vm} and possible lower symmetry $C2/m'$ space groups of generating a magnetic quadrupole at the Cu site in the \HBCO~ unit cell through an ordered arrangement of magnetic dipoles, noting that the arrangement must be antiferromagnetic with no overall magnetization. Within the $Cm'm'm'$ magnetic space group, magnetic dipole moments on atomic sites can be non-zero only on the apical oxygen or barium atoms, and in both cases must be aligned along the tetragonal $c$ axis (Fig.~\ref{fig2} (b)). As mentioned above, symmetry analysis (see Tab.~\ref{tab:mag_symmetry}) shows that the $m'm'm'$ point group symmetry at the Cu site permits either an $M_{z^2}$ or $M_{x^2-y^2}$ quadrupole on the Cu ion but no magnetic dipole moment. We find that $M_{x^2-y^2}$ quadrupoles can not be induced from combinations of symmetry-allowed magnetic dipole moments if the moments are restricted to lie on the atomic sites. Static magnetic moments on Ba and/or O aligned along the $c$ axis, on the other hand, do create a $M_{z^2}$ quadrupole on Cu, as illustrated in Fig.~\ref{fig2} (b). 

We emphasize, however, that the presence of a magnetic moment on either oxygen or barium is unlikely, particularly in the absence of magnetic moments on Cu, due to the closed-shell configurations of both ions in their formal charge state. To investigate the energetics, we performed a constrained-moment density functional calculation with the oxygen magnetic moments set to the pattern of Fig.~\ref{fig2} (b) and the magnitude of \unit[0.1]{\mub} suggested by the neutron data, and find an energy {\it cost} of $\sim$100meV per formula unit over the non-magnetic case. We recall, in addition, that such large magnetic dipole moments are excluded by the absence of internal magnetic fields indicated by NMR measurements, which in the case of \HBCO~ have an upper bound of 0.1mT \cite{Mounce_et_al:2013}.

The lower symmetry $C2/m'$ space group can be reached by canting of the oxygen magnetic moments away from the $z$ axis occurs, as illustrated in Fig.~\ref{fig2} (c). The magnetoelectric multipole illustrated on the Cu atoms in Fig.~\ref{fig2} (c) is the equal linear combination of the $M_{z^2}$ and $M_{xz}$ quadrupoles; other combinations as listed in Tab.~\ref{tab:mag_symmetry} are allowed by symmetry. The difficulties discussed above -- that creating dipole moments on O in the absence of ordered moments on Cu is energetically expensive, and that the absence of internal fields in the NMR data permits only very small values of magnetic dipole moments -- persist for this site symmetry.

\begin{table}[tb]
\caption{Magnetic quadrupole tensor elements, $M_{ij}$, the phonon mode symmetries by which they are induced, the magnetic point groups they create if situated on the Cu ion in Hg-1201 and the non-zero elements of the corresponding magnetoelectric (ME) tensors. In the lower part we list equal and unequal ($\chi\neq1$) superpositions of quadrupoles and the corresponding symmetries.
\label{tab:mag_symmetry} }
\begin{ruledtabular}
\begin{tabular}{|c|c|c|c|}
$M_{ij}$	& phonon &point group 	& ME tensor  	\rule[-1ex]{0pt}{4ex} \\\hline
$z^2$,~$x^2-y^2$			& \PH{A}{2u},\PH{B}{2u}&	$m'm'm'$		&	$\begin{bmatrix}\alpha_{11} &  &  \\ & \alpha_{22} &  \\ &  &  \alpha_{33}\end{bmatrix}$	\rule[3ex]{0pt}{3ex} \\
$xz$/$yz$		& \PH{E}{u}	&	$m'mm$		&	$\begin{bmatrix} \phantom{\alpha_{xx}} &  &  \\ & \phantom{\alpha_{xx}} &\alpha_{23}  \\ &\alpha_{32}  &  \end{bmatrix}$\rule[3ex]{0pt}{4ex} \\
\begin{tabular}{@{}c@{}}$z^2$/$(x^2-y^2)$+ \\ +$xz$/$yz$\end{tabular} 	& \PH{A}{2u}/\PH{B}{2u}+\PH{E}{u}&	$2/m'$	&	$\begin{bmatrix} \alpha_{11} &  & \alpha_{23} \\ & \alpha_{22} &  \\\alpha_{32} &  &  \alpha_{33}\end{bmatrix}$\rule[3ex]{0pt}{4ex} \\
$xz$+$\chi~yz$	&\PH{E}{u}	&	$2'/m$	&	$\begin{bmatrix} & \alpha_{12}  &  \\\alpha_{21}  & & \alpha_{23}  \\&  \alpha_{32} & \end{bmatrix}$\rule[3ex]{0pt}{4ex} \\
$xz$+$\chi~yz$+$z^2$	&\PH{E}{u}+	\PH{A}{2u}&	$-1'$	&	$\begin{bmatrix}\alpha_{11} & \alpha_{12}  & \alpha_{13} \\\alpha_{21}  &\alpha_{22} & \alpha_{23}  \\\alpha_{31}&  \alpha_{32} &\alpha_{33} \end{bmatrix}$	\rule[3ex]{0pt}{4ex} \\
\end{tabular}
\end{ruledtabular}
\end{table}

\subsection{Magnetoelectric multipoles from orbital currents}
\begin{figure}[tb]
\includegraphics[width=1\columnwidth]{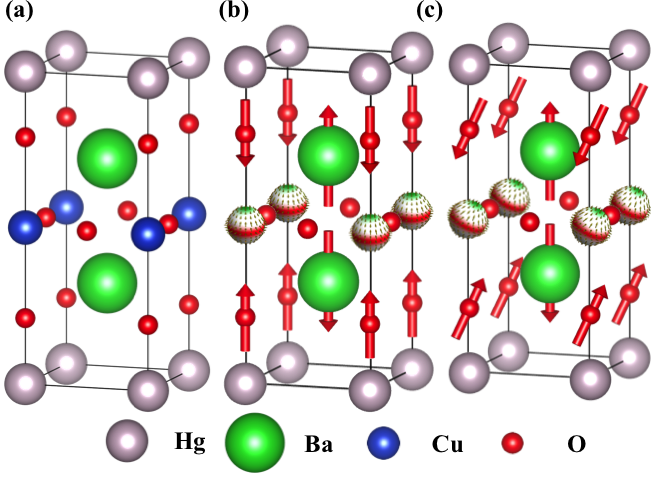}
\caption{\label{fig2} (a) Structure of \HBCO, with Cu, O, Ba, Hg represented by blue, red, green and white spheres, respectively. (b) Allowed dipolar magnetic order for \HBCO within the $Cm'm'm'$ magnetic space group. The red arrows indicate the atomic-site magnetic dipole moments which must be oriented along the $z$ axis. This pattern results in a unit-cell $M_{z^2}$ magnetic quadrupole as well as an atomic-site quadrupole on the Cu site as shown. (c) Possible dipolar magnetic ordering in the $C2/m'$ space group. The dipole moments on the apical oxygens exhibit an additional canting with respect to $Cm'm'm'$. This pattern results in a combination of $M_{z^2}$ and $M_{xz}$ / $M_{yz}$ quadrupoles on the Cu site.}
\end{figure}

In the orbital current mechanism \cite{Varma:1997}, the symmetry considerations discussed above are less restrictive since the magnetic dipole moments generated by the currents are not required to reside on atoms. First we note that the original proposal of currents orbiting within the Cu-O plane (the so-called $\theta_2$ phase) \cite{Varma:1997}, which would generate dipole moments parallel to the $c$ axis, is inconsistent with $Cm'm'm'$ or $C2/m'$ symmetry, which forbids dipole moments parallel to the $c$-axis within the Cu-O plane. In fact the $\theta_2$ phase can be considered to be a combination of a $\vec{T}_y$ toroidal moment with an $xz$ magnetic quadrupole \cite{DiMatteo/Norman:2012}, which we see from Tab.~\ref{tab:mag_symmetry} has $m'mm$ symmetry. Ref.~\onlinecite{Yakovenko:2015} contains a comprehensive discussion of modified orbital current models and proposes one pattern (Fig. 2d of Ref.~\onlinecite{Yakovenko:2015}) that is consistent with all of the latest neutron and Kerr effect data. This orbital current pattern is consistent with our Fig.~\ref{fig2} (c) and corresponds to a combination of $M_{z^2}$ and $M_{xz}$ quadrupoles. Indeed there is a clear correspondence between magnetoelectric multipoles and orbital currents, with the important difference that the orbital currents generate their multipolar character from combinations of magnetic dipoles. We return later to a discussion of the consequences of this difference in microscopic origin, and in particular how it could be used to distinguish the microscopic mechanisms experimentally. 

\subsection{Dynamically ordered magnetoelectric multipoles from spin-phonon coupling}
\begin{figure}[bt]
\includegraphics[width=1\columnwidth]{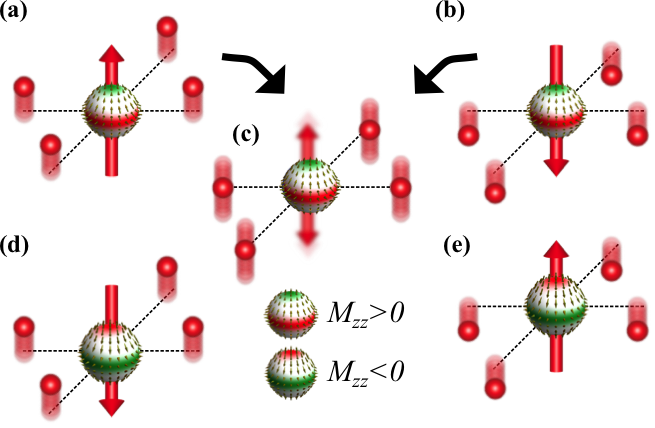}
\caption{\label{fig3}(a-e) Schematics illustrating the formation of a quasi-static magnetic quadrupole via the spin-phonon coupling mechanism. The red spheres represent oxygen ions, which form a square-planar coordination around the central Cu ion. A polar distortion as shown in (a) generates a magnetic quadrupole moment on the central Cu ion because of its magnetic dipole moment. The magnitude of the magnetic quadrupole moment is determined by the product of the size of the distortion and the size of the dipole moment; its sign is determined by the direction of the distortion and the orientation of the magnetic dipole moment. Therefore when the orientation of the Cu magnetic moment changes (d) without a change in the direction of distortion, the opposite magnetic quadrupole moment is generated; likewise when the direction of distortion changes (e) without a change in the magnetic moment orientation, the magnetic quadrupole reverses. If the spin-phonon coupling correlates the orientation of the magnetic dipole moment with the direction of the polar distortion so that both change simultaneously (b) then the sign of the magnetoelectric moment is unchanged (c).}
\end{figure}

Next we introduce a mechanism in which a spontaneous ferroic ordering of magnetic quadrupoles emerges from a dynamically fluctuating system of paramagnetic spin dipole moments on the Cu ions through coupling between the spins and an optical phonon. Within our model, the space-inversion symmetry at the Cu site is broken due to excitation of an optical phonon which shifts the Cu ion from the center of its oxygen coordination plane. Since the Cu ion carries a local (albeit fluctuating) magnetic dipole moment, it therefore acquires a magnetoelectric multipole (again fluctuating) when the space-inversion symmetry is broken. We illustrate this in Fig.~\ref{fig3}  for the case of the $z^2$ magnetic quadrupole coupled to a phonon of $A_{2u}$ symmetry, noting that the mechanism is general, with different magnetoelectric multipoles coupling to different optical phonons as listed in Tab.~\ref{tab:mag_symmetry}.

For a specific direction of off-centering of the Cu ion relative to the oxygen plane, a reversal of the magnetic dipole moment, such as occurs through thermal fluctuation, simultaneously reverses the magnetic quadrupole (Fig.~\ref{fig3} (a) and (b)). Therefore, as expected, a straightforward Langevin disordered-local-moment paramagnet also has no net ordering of its magnetic quadrupoles. Note, however, that a reversal of the direction of displacement of the Cu ion relative tot he oxygen ions, without a reversal of the magnetic dipole moment, also reverses the local Cu quadrupole (Fig.~\ref{fig3} (e)).  Therefore, if a coupling between the local moment and the optical phonon exists, such that a reversal of the local dipole moment is accompanied by a reversal of the Cu ionic displacement, the net quadrupole moment is non-zero. Within this picture, while the time averages of both the magnetic dipole moment, $\braket{\vec{\mu}}_T$, and and the atomic displacements, $\braket{\vec{\vec{r}}}_T$ are zero, the time-averaged expectation values of the form $\braket{\vec{\mu}\cdot\vec{r}}_T$ in particular that of the magnetic quadrupole, are non-zero (Fig.~\ref{fig3} (c)). 

The magnitude of this quasi-static quadrupole within the unit cell is determined by the distance that the Cu ion is displaced from its centrosymmetric position, $r_z$, and is given by $M_{z^2}= \frac{2}{3} r_z \mu_z$, where $\mu_z$ is the magnitude of the fluctuating magnetic dipole moment on the Cu ion ($\sim$1$\mub$ for $d^9$ Cu). For a Cu displacement of \unit[5]{pm}, which might be expected in a thermally activated phonon at the pseudo-gap ordering temperature, the magnitude of the unit-cell quadrupolization is around two orders of magnitude smaller than that of the prototypical magnetoelectric Cr$_2$O$_3$. In addition, the Cu ion develops a local on-site quadrupole due to the local asymmetry in its magnetization density as discussed above; we will calculate its magnitude in the next section.

\section{Computational Details}

Our electronic structure calculations are based on density functional theory within the local spin density approximation (LSDA). To account for the strong electron-electron interactions on the Cu $d$ orbitals, we incorporate the widely used Hubbard $U$ and $J$ corrections of \unit[8]{eV} and \unit[1]{eV} respectively \cite{Anisimov:1991wt,Anisimov/Aryasetiawan/Liechtenstein:1997,Wu:2006gj,Himmetoglu:2011gn} within the LSDA$+U$ method, with double counting treated within the fully-localized limit. Structural parameters, phonon frequencies and spin-phonon coupling constants are calculated using the Vienna \textit{ab-initio} simulation package (VASP) \cite{Kresse:1996vf} within the projector augmented wave (PAW) method \cite{Blochl:1994uk} using default VASP PAW pseudopotentials with the following electrons in the valence: Hg ($5d^{10}6s^2$), Ba ($5s^25p^66s^2$), Cu ($3p^64s^13d^{10}$) and O ($2s^23p^4$). Convergence of forces to \unit[0.01]{meV/\AA} are obtained with a 17$\times$17$\times$11 $k$-point mesh in combination with a cutoff energy of 600 eV.
 
We calculate the atomic-site magnetoelectric multipoles via a generalized density matrix decomposition using our recent implementation \cite{Spaldin_et_al:2013} within the linearized augmented plane wave (LAPW) \textit{ELK} code \cite{ELK}. The unit-cell quadrupoles are calculated according to Eqn.~\eqref{eq:quad} by summation over local moments, taken to be  the projected magnetic dipole moments in each muffin-tin sphere. As numerical parameters for the LAPW calculations we use a basis set of $l_{max(apw)}$=12, the same $k$-point sampling of the Brillouin zone as in the VASP calculations, and we take the product of the muffin-tin radius (1.2, 1.5, 1.1 and 0.7 \AA~for  Hg, Ba, Cu and O, respectively) and the maximum reciprocal lattice vector to be 7.5. The total energy for both codes is converged to within \unit[1]{$\mu$eV} using these settings. To introduce hole doping we reduce the total electron count in our self-consistent calculation by the corresponding amount, while adding a uniform background charge to maintain charge neutrality, as used previously in Refs.~\onlinecite{AmbroschDraxl:2003ff,Thonhauser:2004jr,AmbroschDraxl:2004kp,AmbroschDraxl:2004bt,AmbroschDraxl:2006jc} to calculate various doping-dependent properties of the \HBCOx~family.

\section{Density functional calculations of spin-phonon coupling and magnetoelectric multipoles in H\lowercase{g}-1201}

We begin by calculating the structure of $\delta=0.05$ doped Hg-1201 without allowing for spin polarization to obtain a paramagnetic reference state. Our resulting structure (Tab.~\ref{tab:struct}) agrees well with the experimentally reported structure (Ref.~[\onlinecite{Huang:1995tn}]) while showing the usual LDA underestimation of the unit cell volume (\unit[3.8]{\%} in this case). Using our calculated structure we then compute the total energies and phonon spectra, via the frozen phonon method, for ferromagnetic and checkerboard antiferromagnetic (in a $\sqrt{2}\times\sqrt{2}\times 1$ supercell) orderings. 

\begin{table}[bt]
\caption{\label{tab:struct} Experimental (EXP) \cite{Huang:1995tn} and calculated in this work (DFT) lattice constants and atomic positions for \HBCO. The Wykoff positions are given according to the $Cm'm'm'$ magnetic space group; 4l and 4k are the only internal degrees of freedom.}
\begin{ruledtabular}
\begin{tabular}{|lcll|}
\multicolumn{2}{l}{lattice constant} & DFT & EXP  \rule[-1ex]{0pt}{3.5ex} \\\hline
a [\AA]		&   &3.84	&3.87	\rule[-1ex]{0pt}{3.5ex} \\
c/a   		&   & 2.44	&2.45	\rule[-1ex]{0pt}{3.5ex} \\\hline
atom& Wykoff position	& DFT & EXP  \rule[-1ex]{0pt}{3.5ex} \\\hline
Hg		& 2a&0.00 	& 0.00 	\rule[-1ex]{0pt}{3.5ex}\\
Ba		& 4l &0.30	& 0.30	\rule[-1ex]{0pt}{3.5ex}\\
Cu		& 2b &0.50	& 0.50	\rule[-1ex]{0pt}{3.5ex}\\
O$_1$	& 4f &0.50	& 0.50	\rule[-1ex]{0pt}{3.5ex}\\
O$_2$	& 4k &0.21	& 0.21	\rule[-1ex]{0pt}{3.5ex}\\
\end{tabular}
\end{ruledtabular}
\end{table}

The nearest-neighbor Heisenberg exchange constant, $J$, is then obtained directly from the energy difference between the ferromagnetically ordered and in-plane checkerboard antiferromagnetically ordered magnetic moment arrangements within the same crystal structure, $J=-\frac{1}{2}(E_{\textnormal{AFM}}-E_{\textnormal{FM}})$. We obtain $J=$\unit[78]{meV}, consistent with literature values \cite{Anonymous:ZSMerAaE}. Here $E_{\textnormal{AFM}}$ is the energy of the $\sqrt{2}\times\sqrt{2}\times 1$ AFM-ordered unit cell containing two formula units, and $E_{\textnormal{FM}}$ is the corresponding FM energy. We take as phonon frequencies $\omega_i=\sqrt{\frac{\omega^2_{\textnormal{FM}_i}+\omega^2_{\textnormal{AFM}_i}}{2}}$, where $\omega_{\textnormal{AFM}_i}$ and $\omega_{\textnormal{FM}_i}$ are the frequencies of mode $i$ for the AFM and FM magnetic orderings respectively. In Tab.~\ref{tab:phonon} we compare our calculated phonon frequencies with available experimental data \cite{dAstuto:2003cy,Uchiyama:2004kl} and obtain excellent agreement, with the largest deviation from experiment being only \unit[6]{\%}. 

\begin{figure}[bt]
\includegraphics[width=1\columnwidth]{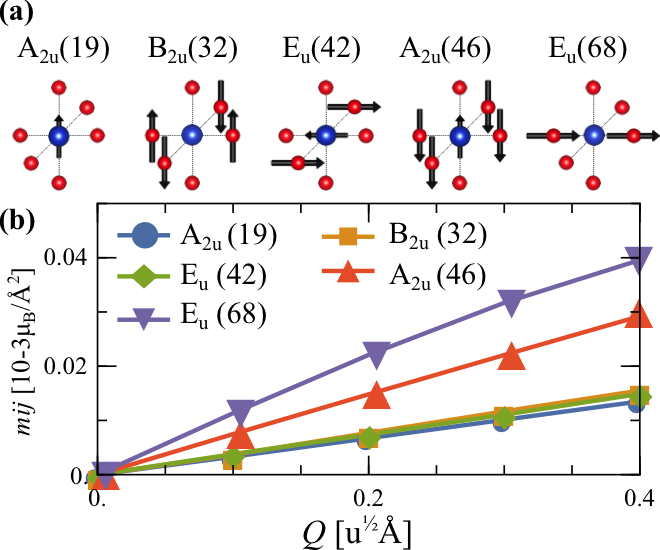}
\caption{\label{fig4}(a) Schematics of CuO modes with the strongest spin-phonon coupling. The labels above the pictures represent the mode symmetry and frequency in meV. (b) Variation of the local quadrupolar moments $m_{ij}$ as a function of phonon mode amplitudes for all modes shown schematically in (a).}
\end{figure}

The lowest order spin-phonon coupling between a polar phonon and a parity-even spin arrangement is quadratic in both spin and phonon amplitude, therefore we extract the spin-phonon coupling constants as 
\begin{equation}
g_i=\frac{\omega^2_{\textnormal{FM}_i}-\omega^2_{\textnormal{AFM}_i}}{4~S^2_l} \quad ,
\end{equation}
where $S_l$ is the size of the local spin dipole moment.
We find that, of the nine non-translational polar modes, only seven have a sizable frequency difference between the FM and AFM orderings, and hence a considerable spin-phonon coupling. All of these strongly spin-phonon coupled modes, which we show in Fig.~\ref{fig4} (note that two are two-fold degenerate) correspond to relative displacements of the Cu cation and its coordinating oxygen anions. The frequencies of these modes span the range from \unit[20-70]{meV} and we list the corresponding couplings in Tab.~\ref{tab:coupling}. (In the following we use as labels for each phonon mode its irreducible representation followed in brackets by its frequency in meV.) 

\begin{table}[bt]
\caption{Comparison of experimental (EXP) and calculated in this work (DFT) phonon frequencies at the zone center ($q=0$) of Hg-1201. The values marked with a star ($^*$) were obtained from a shell model calculation and an interpolation to the zone center.}\label{tab:phonon}
\begin{ruledtabular}
\begin{tabular}{|c|c|c|c|}
mode	        &    \multicolumn{3}{c}{frequency, $\omega$ [meV]}  \rule[-1ex]{0pt}{3.5ex}\\
symmetry	&  DFT & EXP\cite{Uchiyama:2004kl} & EXP\cite{dAstuto:2003cy}       \rule[-1ex]{0pt}{3.5ex}\\\hline
\PH{E}{u\phantom{u}}		&	\phantom{1}7.7	&\phantom{1}7.24$^*$	& 		\rule[-1ex]{0pt}{3.5ex}\\
\PH{E}{g\phantom{u}}		&	\phantom{1}9.3	&\phantom{1}9.43\phantom{$^*$}	& 		\rule[-1ex]{0pt}{3.5ex}\\
\PH{A}{2u}		&	10.3	&		&		\rule[-1ex]{0pt}{3.5ex}\\
\PH{E}{u\phantom{u}}		&	19.0	&19.9$^*$	&		\rule[-1ex]{0pt}{3.5ex}\\ 
\PH{A}{2u}		&	19.4	&		&		\rule[-1ex]{0pt}{3.5ex}\\
\PH{A}{1g}		&	20.2	&20.1$^*$	&20.0	\rule[-1ex]{0pt}{3.5ex}\\
\PH{E}{g\phantom{u}}		&	21.2	&		&20.5	\rule[-1ex]{0pt}{3.5ex}\\
\PH{E}{u\phantom{u}}		&	28.9	&29.5$^*$	&		\rule[-1ex]{0pt}{3.5ex}\\
\PH{B}{2u}		&	32.0	&		&		\rule[-1ex]{0pt}{3.5ex}\\
\PH{E}{u\phantom{u}}		&	42.3	&45.4$^*$	&		\rule[-1ex]{0pt}{3.5ex}\\
\PH{A}{2u}		&	46.3	&		&		\rule[-1ex]{0pt}{3.5ex}\\
\PH{E}{u\phantom{u}}		&	68.5	&71.8$^*$	&		\rule[-1ex]{0pt}{3.5ex}\\
\PH{A}{1g}		&	70.8	&72.3$^*$	&73.4	\rule[-1ex]{0pt}{3.5ex}\\
\PH{A}{2u}		&	76.1	&		&		\rule[-1ex]{0pt}{3.5ex}\\
\end{tabular}
\end{ruledtabular}
\end{table}

Freezing in the displacement patterns of each of these polar modes using FM magnetic ordering induces an atomic-site magnetic quadrupole on the Cu site of symmetry determined by the symmetry of the phonon: $M_{z^2}$ for the \PH{A}{2u} modes, $M_{x^2-y^2}$ for the \PH{B}{2u} mode, and $M_{xz}$ or $M_{yz}$ for the \PH{E}{u} modes, where for the latter case the orientation of the doubly degenerate mode determines the symmetry of the quadrupole. In Fig.~\ref{fig4} we show our calculated magnitudes for the Cu atomic-site magnetic quadrupoles, obtained from decomposition of the density matrix \cite{Spaldin_et_al:2013}, as a function of the amplitude of the phonon eigenvector for all five independent modes. We see that the response is linear in each case, with the \PH{E}{u}(68) and \PH{A}{2u}(46) modes exhibiting the largest slope. To be able to compare the induced moments we list the atomic-site magnetic quadrupole moments, $m_{ij}$, at one quarter amplitude of the normalized eigenvector ($Q=0.25$) for each mode in Tab.~\ref{tab:coupling}. While the amplitude of the phonon excitations depends on both the frequency and the temperature, a quarter amplitude corresponds approximately to the average of their thermally excited displacements at \unit[200]{K} and represents a maximum relative copper-oxygen displacement of 6pm. Note that the quadrupolar moment in the undistorted structure is zero by symmetry. 

Next, we calculate the induced unit-cell magnetic quadrupole, $M_{ij}$, for each mode using Eqn.~\eqref{eq:quad}, where the integral is replaced by the sum over the local dipole moments $\mu$ on the Cu ions and the positions of the local dipole moments are set by the mode pattern of each phonon mode. Note that the symmetries of the allowed induced unit-cell quadrupoles are the same as those of the atomic-site quadrupoles for the corresponding mode. As before, the magnitude of the induced quadrupole is linear in the amplitude of the mode; in Tab.~\ref{tab:coupling} we list the $M_{ij}$ values again for $Q=0.25$. We find the strongest unit-cell quadrupolar responses from the \PH{A}{2u}(46) and \PH{A}{2u}(19) modes, both of which exhibit a significant amount of Cu movement (see Fig.~\ref{fig3} (a)). In contrast, the asymmetric mode pattern of \PH{B}{2u}(32) induces no unit-cell magnetic quadrupole on symmetry grounds.

The total quadrupole moment per unit cell induced by a specific mode is the sum of atomic-site and unit-cell contributions. We find for most modes that the latter contribution is dominant and at least one order of magnitude larger than the atomic-site quadrupole. Consequently, the \PH{A}{2u}(46) and \PH{A}{2u}(19) modes induce the largest magnetic quadrupoles in Hg-1201. Our quantitative analysis confirms our earlier estimate that the size of these magnetic quadrupoles for typical thermally activated mode amplitudes is approximately two orders of magnitude smaller than that of Cr$_2$O$_3$. 

Comparison of the magnitudes of the spin-phonon coupling and the induced magnetic quadrupoles shows that the strongest spin-phonon coupling does not necessarily result in the largest magnetic quadrupoles; specifically the strongest spin-phonon coupling is found for the \PH{E}{u}(42) mode, whereas the \PH{E}{u}(68) mode induces the strongest atomic-site and the \PH{A}{2u}(19) mode the strongest unit-cell magnetoelectric quadrupolar moment. The spin-phonon coupling is determined by the change in magnetic exchange interaction on changing bond angles and lengths, whereas the source of the induced unit-cell quadrupolar moment is the generated asymmetry in magnetization at the Cu site. The exchange interactions are dominated in turn by the Cu-O-Cu bond angle, which changes by \unit[8.2]{$^{\circ}$} per mode amplitude for \PH{E}{u}(42) and only \unit[2.8]{$^{\circ}$} for \PH{A}{2u}(19). In contrast the spin asymmetry also depends on the shift of the Cu ion away from its centrosymmetric position; this is \unit[0.10]{pm} per mode amplitude for \PH{A}{2u}(19) and only \unit[0.04]{pm} for \PH{E}{u}(42). 

\section{Spin-phonon coupling and dynamical ferro-ordering of the magnetoelectric multipoles}

Finally, we present a simple ``toy model'' simulation of the time evolution of our coupled spin-phonon system to illustrate that, while the spin fluctuations yield an average magnetization of zero, and the phonon vibrations yield an average lattice displacement of zero, the coupling between the two yields a non-zero average magnetoelectric multipole. We use the following Hamiltonian to describe the coupling between the polar phonon modes and the spin-lattice: 
\begin{equation}\label{eq_spinphonon}
\hat{H}=J\sum\limits_{\braket{nn}} S_l\cdot S_m +\sum\limits_i \frac{\omega_i^2}{2}Q_i^2 +\sum\limits_{\braket{nn},i}g_i  Q_i^2   S_l\cdot S_m \quad.
\end{equation}
Here $\sum_{\braket{nn}}$ and $\sum_{i}$ indicate summation over nearest-neighbor spins and phonon modes, respectively. The first term is the usual Heisenberg spin Hamiltonian in which localized spins $S_l$ interact via exchange interaction $J$, and the second term is the usual harmonic oscillator, with $Q_i$ giving the amplitude of the (in this case polar) $i$th phonon mode of frequency $\omega_i$. The last term describes the lowest order spin-phonon coupling between polar modes and parity-even spin arrangements, with coupling constant $g_i$. We use the values of the magnetic exchange constant $J$, phonon frequencies $\omega_i$ and spin-phonon coupling constants $g_i$ calculated in the previous section using density functional theory for the composition HgBa$_2$CuO$_{4.05}$ ($\delta=0.05$). Note that coupling to the phonon modulates the exchange interaction between the spins creating an effective time-dependent magnetic exchange constant $J_{\textnormal{eff}}=J+g_i~Q_i(t)^2$. Likewise, coupling to the spin system modulates the phonon frequency, giving $w_{\textnormal{eff},i}=2\sqrt{w_i^2/2+g_i\sum S_l\cdot S_m}$. 

\begin{table}[tb]
\caption{\label{tab:coupling} Phonon frequencies, spin-phonon coupling constants, magnetic quadrupole symmetry and atomic-site, $m_{ij}$, and unit-cell,
$M_{ij}$, magnetic quadrupole values for quarter amplitudes ($Q=0.25$) of selected polar modes in Hg-1201.}
\begin{ruledtabular}
\begin{tabular}{|c|c|c|c|c|c|}
mode 	&$\omega$  	& g  			& quadrupole			& $m_{ij}$	&	$M_{ij}$\rule[-1ex]{0pt}{3.5ex} \\
symmetry	 &[meV]& $\left[\textnormal{meV}\right]$&symmetry &  $\left[10^{-3}\frac{\mub}{\textnormal{\AA}^2}\right]$ & $\left[10^{-3}\frac{\mub}{\textnormal{\AA}^2}\right]$\rule[-1ex]{0pt}{3.5ex} \\\hline
\PH{A}{2u}  	&19.4		& \phantom{0}-0.8&$z^2$&0.01 &   0.70\rule[-1ex]{0pt}{3.5ex} \\
\PH{B}{2u}  	&32.0		& \phantom{0}-8.1&$x^2-y^2$	&0.01 &   0\rule[-1ex]{0pt}{3.5ex} \\
\PH{E}{u}  	&42.3		& -14.7			 &$xz/yz$  &0.01 &   0.10\rule[-1ex]{0pt}{3.5ex} \\
\PH{A}{2u}  	&46.3		& \phantom{0}-5.7&$z^2$	&0.02 &   0.41\rule[-1ex]{0pt}{3.5ex} \\
\PH{E}{u}  	&68.5		&\phantom{0}-7.0 &$xz/yz$	&0.03 & 0.00\rule[-1ex]{0pt}{3.5ex} \\
\end{tabular}
\end{ruledtabular}
\end{table}

The spin-phonon coupling contributes a force on the atoms, through $F=-d\hat{H}/dQ=(\omega^2+g \sum\limits_{\braket{n}}S_l\cdot S_m)Q$. We treat this spin-phonon term as a driving force on the oscillator and solve the classical equation of motion 
\begin{equation}\label{eq_dynamics}
Q''(t)+2\gamma Q'(t)+\omega_0^2Q=2g~Q~S_{lm}(t) \quad,
\end{equation}
where $S_{lm}(t)=\sum_{\braket{nn}}S_l(t)\cdot S_m(t)$ is the time-dependent exchange sum and we include a Stokes friction term, $\gamma$, to account for the finite lifetimes of oscillator modes. 

\begin{figure}[bt]
\includegraphics[width=1\columnwidth]{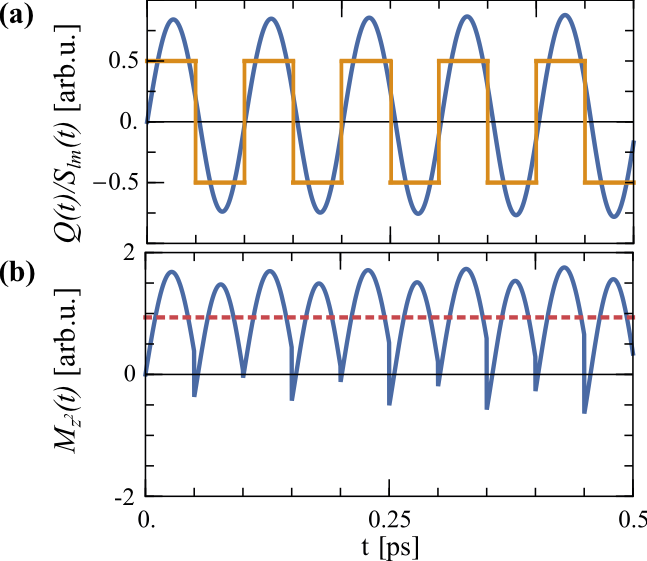}
\caption{\label{fig5}(a) Phonon $Q(t)$ for the mode \PH{A}{2u}(46) (blue) and spin $S_{lm}(t)$ (orange) amplitude as a function of time. (b) Expectation value of the quadrupole moment $M_{z^2}(t)$ as a function of time. The red dotted line shows the time average.}
\end{figure}

In Fig.~\ref{fig5} (a) and (b) we show our calculated time dependences of the atomic displacements, spin expectation value and magnetic quadrupole moment respectively, obtained by solving Eqn.~\eqref{eq_dynamics} \cite{numerical}. Our calculations were performed for parameters of the \PH{A}{2u}(46) mode, and with the spin term taken to be $S_{lm}\propto sgn[\sin(\omega t)]$, with the $sgn$ function quantizing the spin fluctuations to adopt values $S_l=\pm 1/2$. The spin-fluctuation frequency, $\omega$, was set to $\omega_0=$\unit[46]{meV}, the eigenfrequency of the \PH{A}{2u} mode. We see that, as expected, the time-averaged expectation values of both the mode amplitude and the spin expectation values, $\braket{Q}_T$ and $\braket{S_{lm}(t)}_T$, are zero, however the expectation value of the magnetoelectric multipole, $\braket{Q\cdot S_{lm}(t)}_T$ is non-zero. Consequently, a quasi-static magnetoelectric multipolar order arises from the spin-phonon coupling even in the absence of static magnetic and electric dipolar order. When we shift the frequency of the spin flucutations away from $\omega_0$, the maximum and average magnetic quadrupolar amplitudes decrease as the driving term frequency becomes non-resonant with the mode.

\section{Comparison with experiments}

Next we compare the behavior of our proposed quasi-static magnetoelectric quadrupolar order in Hg-1201 with experimental measurements of the behavior of the pseudo-gap phase. First, we note that that any of the \PH{A}{2u} or \PH{B}{2u} modes, which induce $M_{z^2}$ or $M_{x^2-y^2}$ magnetic quadrupoles respectively, result in the $m'm'm'$ Cu point group symmetry that is consistent with existing neutron diffraction measurements. Of these, the \PH{A}{2u}(46) is compatible with the observed Ising-like excitation \cite{Li:2012is}, and combines strong spin-phonon coupling with a sizable quadrupolar response, therefore we restrict any mode-specific discussion to the \PH{A}{2u}(46) mode for conciseness. We note, however, that the model is more generally applicable, and for example the possible lower symmetry $2/m'$ symmetry can be reached from a combination of the \PH{A}{2u} or \PH{B}{2u} modes, and the \PH{E}{u} modes, which induce $M_{xz}$ or $M_{yz}$ quadrupoles and a $T_y$ toroidal moment. We note also that most high-$T_c$ cuprates show similar polar phonons within the copper-oxygen plane at similar frequencies, and so we do not limit our comparison to experimental data for Hg-1201.

\begin{figure}[tb]
\includegraphics[width=1\columnwidth]{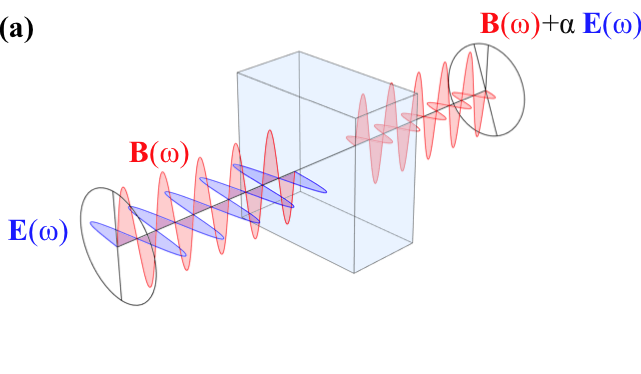}
\caption{Mechanism for the Kerr effect from a diagonal magnetoelectric material. The electromagnetic wave
induces a magnetization parallel to its electrical component through the magnetoelectric effect. The net 
magnetization, and hence the plane of polarization of the light, therefore rotates relative to that of the incoming wave. \label{Kerr}}
\end{figure}

We begin with a discussion of the Kerr effect data. Ref.~\onlinecite{Orenstein:2011} showed that a Kerr rotation should be induced by a magnetoelectric material with either the $m'm'm'$ or $2/m'$ magnetic point group symmetries that form the basis of our model. The mechanism for the Kerr rotation rests on the diagonal linear magnetoelectric response, as illustrated by the cartoon in Fig.~\ref{Kerr}. The strength of the Kerr rotation is determined in part by the strength of the magnetoelectric coupling, for which the magnetic quadrupole provides the measure in our model. Therefore we anticipate a Kerr rotation that is around two orders of magnitude smaller than that of Cr$_2$O$_3$, which is indeed the case \cite{Krichevtsov:1993gs,Xia:2008cc}. Our quasi-static magnetoelectric multipole model is therefore consistent both qualitatively and quantitatively with the observed Kerr rotation in the pseudo-gap phase. 

Next we discuss the local probe NMR and $\mu$SR measurements. Both methods have been used extensively to search for static magnetic dipole moments in cuprates, motivated in large part by an attempt to verify the orbital-currents model. Neither method, applied to any of the cuprates, have succeeded in finding internal magnetic fields corresponding to a static dipolar ordering, giving vanishingly small upper bounds on the possible static dipole moments \cite{MacDougall:2008br,Strassle:2008ku,Strassle:2011dr}. Here we calculate the internal fields, from point quadrupolar calculations, generated by the ferroic ordering of the local, atomic-site magnetic quadrupoles. We find that the local $M_{z^2}$ quadrupole on Cu which is induced by quarter displacement of the \PH{A}{2u}(46) mode generates a magnetic field of only \unit[0.03]{mT} at its neighboring oxygen sites. This field strength is below the static detection limit of NMR and so is consistent with the absence of a signal in these measurements. We note also that fluctuating dipolar moments at frequencies above \unit[0.04]{GHz} ($\mu$SR) or \unit[100]{GHz} (NMR), could not be excluded from the resonance experiments. In our mechanism the spin oscillates at frequencies close to the eigenfrequency of the \PH{A}{2u}(46) phonon mode, which is in the THz range and thus significantly above the detection limit of the resonance measurements. 

While the compatibility of the symmetry of ferroic ordering of Cu $M_{z^2}$ magnetic quadrupoles with neutron measurements has already been proposed \cite{Lovesey:2014vm,Lovesey:2015wb}, here we show that such magnetoelectric multipoles also resolve the paradox between the values of local magnetic dipole moments inferred from interpreting the neutron data in terms of magnetic dipole scattering, and the much smaller values measured from the local probe NMR and $\mu$SR methods. Detailed analysis of the neutron scattering data for YBCO showed that the signal could be interpreted in terms of magnetic dipole moments on the oxygen atoms with a maximum magnitude of $\sim$0.1$\mub$ \cite{Fauque:2006bo}, apparently incompatible with the much lower limits set by NMR and $\mu$SR. When we perform a DFT calculation with the oxygen dipole moments constrained to this value we generate an atomic-site magnetic quadrupole on Cu that has a magnitude close to that 
generated by our quasi-static mechanism. (Note, however, that, as we stated previously, it is energetically costly to force magnetic dipole moments onto the closed shell oxygen atoms.) We propose, therefore, that the neutron measurements are sensitive directly to the magnetoelectric multipoles, and the apparent inconsistencies between the various experimental methods are an artifact of an attempt to map the magnetoelectric multipolar ordering onto a dipolar picture. 

A defining characteristic of the pseudo-gap phase is, of course, the existence of a pseudo-gap. We searched for this by monitoring the change in the density of states on freezing in the \PH{A}{2u}(46) pattern of atomic displacements with progressively increasing amplitude, and indeed found a reduction in the density of states at the Fermi level, $N(E_F)$. This reduction will in turn reduce the electronic specific heat and conductance, as observed in experiment \cite{Timusk:1999wp}. Since our order parameter is not a conventional static Landau-type order parameter, however, there is no associated soft mode and we do not expect a discontinuity in specific heat. Again this is compatible with experimental measurements.

Perhaps the most compelling evidence for the mechanism proposed here is the recent observation using neutron scattering of an Ising-like excitation at $\sim$\unit[40]{meV} in Hg-1201 at the onset to the pseudo-gap phase \cite{Li:2012is}. We assign this Ising-like excitation to our coupled spin-\PH{A}{2u}(46) phonon mode responsible for the quadrupolar order; our calculations of the doping-dependence of the phonon frequencies indicate that the difference in energy due to the higher doping level, $\delta_{\textnormal{exp}}\approx 0.11$, used in the experiment. Intriguingly, we also find a strong doping dependence of the spin-phonon coupling, with the spin-phonon coupling of the \PH{A}{2u}(46) mode approaching zero as doping is increased beyond the optimally doped range, in the same manner as $T^*$.

\section{Summary and Discussion}

In summary, we have shown that a quasi-static ordering of magnetoelectric multipoles mediated by coupling between paramagnetic spin moments on the Cu ions and a polar lattice phonon, provides a candidate order parameter for the pseudo-gap phase of the high-$T_c$ cuprates. This candidate mechanism is, to our knowledge, consistent with all known experimental data. In addition, it captures the spirit of the earlier orbital-currents model as well as the analysis of neutron scattering in terms of magnetoelectric multipoles, while providing a microscopic origin for the behavior. There remain many open questions regarding this quasi-static magnetoelectric quadrupolar ordering that require further experimental measurement and theoretical analysis:

The first is whether such a dynamically varying ordering that fluctuates with a net non-zero value constitutes an order parameter, and in turn whether its onset should be described as a phase transition. Clearly it does not fit straightforwardly into the usual Landau description of spontaneous symmetry breaking phase transitions. To the extent that it is associated with a well-defined symmetry lowering, it has characteristics of a phase transition. In addition, it has features of a ferroic transition, with multiple equivalent ground states that can in principle form domains. The absence of a soft mode or divergence of the specific heat, however, are clearly unconventional.

Another obvious question is how to measure the proposed effect. One possibility could be a direct mapping of the associated internal quadrupolar magnetic field using $\mu$SR, or the associated field gradients using NMR or NQR; in particular our initial calculations in this direction suggest that these have a strong doping dependence. Such a measurement could also distinguish quasi-static multipoles from those generated by a purely orbital current mechanism, which, as we mentioned earlier, carry also a magnetic dipole contribution. Perhaps a more convincing distinction between the orbital-current mechanism and quasi-static quadrupoles could be revealed from studies of oxygen isotope effects, to which our model should be rather sensitive and the orbital currents not. Measurement of the E1-E2 pre-edge transition with magneto-chiral diffraction would also provide evidence for the quadrupolar order although not for its origin. A successful experiment, however, would require a single domain quadrupolar state. We expect that the lengthscale of the domains should be set by the correlation lengths of the phonon and spin fluctuations, although simultaneous electric and magnetic fields should in principle provide a conjugate field for achieving a single domain state. 

We make no statement at this stage as to whether the quasi-static quadrupolar ordering competes with the superconducting state or is a precursor to it. If the latter, it could harmoniously resolve the ongoing dispute as to whether the pairing mechanism is spin- or lattice-driven since its existence requires a coupling of the two. A possible experiment to address this question would be optical pumping of the associated \PH{A}{2u}(46) phonon mode, in a similar manner as used recently in Refs.~\onlinecite{Hu_et_al:2014} and \onlinecite{Mankowsky_et_al:2014} to study the origin of superconductivity in YBCO. The \unit[46]{meV} frequency is in a particularly inconvenient range in terms of experimental accessibility, however. In the meantime, this intriguing hint of a time- and space- odd order parameter suggests that perhaps the family of multiferroic and magnetoelectric compounds should receive more attention as potential superconductors.

Finally, we note that the model Hamiltonian of Eqn.~\eqref{eq_spinphonon} is not restricted to polar or zone-center modes, and coupling to other modes, or to non-zone-center phonons could give quasi-static ordering of other elusive order parameters, such as charge ordering or stripes, that are associated with exotic superconductivity. 

\section{Acknowledgments}
This work was supported financially by the ETH Z\"urich (MJAF and NAS), by the ERC Advanced Grant program, No. 291151 (MF and NAS), by the Max R\"ossler Prize of the ETH Z\"urich (NAS), and by the Sinergia program of the Swiss National Science Foundation Grant No. CRSII2\_147606/1 (FT, NAS and US). Calculations were performed at the Swiss National Supercomputing Center. We thank Lars Nordstr\"om for assistance with developments in the ELK code, and Bertram Batlogg, Philippe Bourges, Sergio DiMatteo, Pietro Gambardella, Antoine Georges, J\"urgen Haase, Aaron Kapitulnik, Dmitri Khalyavin, Steven Kivelson, Stephen Lovesey, and Jan Zaanen for fruitful discussions and valuable feedback on the manuscript. 

\bibliography{tot,Nicola,NewRefs}

\begin{thebibliography}{55}
\expandafter\ifx\csname natexlab\endcsname\relax\def\natexlab#1{#1}\fi
\expandafter\ifx\csname bibnamefont\endcsname\relax
  \def\bibnamefont#1{#1}\fi
\expandafter\ifx\csname bibfnamefont\endcsname\relax
  \def\bibfnamefont#1{#1}\fi
\expandafter\ifx\csname citenamefont\endcsname\relax
  \def\citenamefont#1{#1}\fi
\expandafter\ifx\csname url\endcsname\relax
  \def\url#1{\texttt{#1}}\fi
\expandafter\ifx\csname urlprefix\endcsname\relax\def\urlprefix{URL }\fi
\providecommand{\bibinfo}[2]{#2}
\providecommand{\eprint}[2][]{\url{#2}}

\bibitem[{\citenamefont{Varma}(2010)}]{Varma:2010}
\bibinfo{author}{\bibfnamefont{C.~M.} \bibnamefont{Varma}},
  \bibinfo{journal}{Nature} \textbf{\bibinfo{volume}{468}},
  \bibinfo{pages}{184} (\bibinfo{year}{2010}).

\bibitem[{\citenamefont{Warren et~al.}(1989)\citenamefont{Warren, Walstedt,
  Brennert, Cava, Tycko, Bell, and Dabbagh}}]{Warren_et_al:1989}
\bibinfo{author}{\bibfnamefont{W.~W.} \bibnamefont{Warren},
  \bibfnamefont{Jr.}}, \bibinfo{author}{\bibfnamefont{R.~E.}
  \bibnamefont{Walstedt}}, \bibinfo{author}{\bibfnamefont{G.~F.}
  \bibnamefont{Brennert}}, \bibinfo{author}{\bibfnamefont{R.~J.}
  \bibnamefont{Cava}}, \bibinfo{author}{\bibfnamefont{R.}~\bibnamefont{Tycko}},
  \bibinfo{author}{\bibfnamefont{R.~F.} \bibnamefont{Bell}}, \bibnamefont{and}
  \bibinfo{author}{\bibfnamefont{G.}~\bibnamefont{Dabbagh}},
  \bibinfo{journal}{Phys. Rev. Lett.} \textbf{\bibinfo{volume}{62}},
  \bibinfo{pages}{1193} (\bibinfo{year}{1989}).

\bibitem[{\citenamefont{Fauqu{\'e} et~al.}(2006)\citenamefont{Fauqu{\'e},
  Sidis, Hinkov, Pailh{\`e}s, Lin, Chaud, and Bourges}}]{Fauque:2006bo}
\bibinfo{author}{\bibfnamefont{B.}~\bibnamefont{Fauqu{\'e}}},
  \bibinfo{author}{\bibfnamefont{Y.}~\bibnamefont{Sidis}},
  \bibinfo{author}{\bibfnamefont{V.}~\bibnamefont{Hinkov}},
  \bibinfo{author}{\bibfnamefont{S.}~\bibnamefont{Pailh{\`e}s}},
  \bibinfo{author}{\bibfnamefont{C.~T.} \bibnamefont{Lin}},
  \bibinfo{author}{\bibfnamefont{X.}~\bibnamefont{Chaud}}, \bibnamefont{and}
  \bibinfo{author}{\bibfnamefont{P.}~\bibnamefont{Bourges}},
  \bibinfo{journal}{Phys. Rev. Lett.} \textbf{\bibinfo{volume}{96}},
  \bibinfo{pages}{197001} (\bibinfo{year}{2006}).

\bibitem[{\citenamefont{Shekhter et~al.}(2013)\citenamefont{Shekhter, Ramshaw,
  Liang, Hardy, Bonn, Balakirev, McDonald, Betts, Riggs, and
  Migliori}}]{Shekhter_et_al:2013}
\bibinfo{author}{\bibfnamefont{A.}~\bibnamefont{Shekhter}},
  \bibinfo{author}{\bibfnamefont{B.~J.} \bibnamefont{Ramshaw}},
  \bibinfo{author}{\bibfnamefont{R.}~\bibnamefont{Liang}},
  \bibinfo{author}{\bibfnamefont{W.~N.} \bibnamefont{Hardy}},
  \bibinfo{author}{\bibfnamefont{D.~A.} \bibnamefont{Bonn}},
  \bibinfo{author}{\bibfnamefont{F.~F.} \bibnamefont{Balakirev}},
  \bibinfo{author}{\bibfnamefont{R.~D.} \bibnamefont{McDonald}},
  \bibinfo{author}{\bibfnamefont{J.~B.} \bibnamefont{Betts}},
  \bibinfo{author}{\bibfnamefont{S.~C.} \bibnamefont{Riggs}}, \bibnamefont{and}
  \bibinfo{author}{\bibfnamefont{A.}~\bibnamefont{Migliori}},
  \bibinfo{journal}{Nature} \textbf{\bibinfo{volume}{498}}, \bibinfo{pages}{75}
  (\bibinfo{year}{2013}).

\bibitem[{\citenamefont{Straessle et~al.}(2011)\citenamefont{Straessle,
  Graneli, Mali, Roos, and Keller}}]{Strassle:2011dr}
\bibinfo{author}{\bibfnamefont{S.}~\bibnamefont{Straessle}},
  \bibinfo{author}{\bibfnamefont{B.}~\bibnamefont{Graneli}},
  \bibinfo{author}{\bibfnamefont{M.}~\bibnamefont{Mali}},
  \bibinfo{author}{\bibfnamefont{J.}~\bibnamefont{Roos}}, \bibnamefont{and}
  \bibinfo{author}{\bibfnamefont{H.}~\bibnamefont{Keller}},
  \bibinfo{journal}{Phys. Rev. Lett.} \textbf{\bibinfo{volume}{106}},
  \bibinfo{pages}{097003} (\bibinfo{year}{2011}).

\bibitem[{\citenamefont{Tokunaga et~al.}(1999)\citenamefont{Tokunaga, Ishida,
  Magishi, Ohsugi, Zheng, Kitaoka, Asayama, Iyo, Tokiwa, and
  Ihara}}]{Tokunaga:1999df}
\bibinfo{author}{\bibfnamefont{Y.}~\bibnamefont{Tokunaga}},
  \bibinfo{author}{\bibfnamefont{K.}~\bibnamefont{Ishida}},
  \bibinfo{author}{\bibfnamefont{K.}~\bibnamefont{Magishi}},
  \bibinfo{author}{\bibfnamefont{S.}~\bibnamefont{Ohsugi}},
  \bibinfo{author}{\bibfnamefont{G.~q.} \bibnamefont{Zheng}},
  \bibinfo{author}{\bibfnamefont{Y.}~\bibnamefont{Kitaoka}},
  \bibinfo{author}{\bibfnamefont{K.}~\bibnamefont{Asayama}},
  \bibinfo{author}{\bibfnamefont{A.}~\bibnamefont{Iyo}},
  \bibinfo{author}{\bibfnamefont{K.}~\bibnamefont{Tokiwa}}, \bibnamefont{and}
  \bibinfo{author}{\bibfnamefont{H.}~\bibnamefont{Ihara}},
  \bibinfo{journal}{Physica B: Condensed Matter}
  \textbf{\bibinfo{volume}{259-261}}, \bibinfo{pages}{571}
  (\bibinfo{year}{1999}).

\bibitem[{\citenamefont{Sonier et~al.}(2001)\citenamefont{Sonier, Brewer,
  Kiefl, Miller, and Morris}}]{Sonier_et_al:2001}
\bibinfo{author}{\bibfnamefont{J.~E.} \bibnamefont{Sonier}},
  \bibinfo{author}{\bibfnamefont{J.~H.} \bibnamefont{Brewer}},
  \bibinfo{author}{\bibfnamefont{R.~F.} \bibnamefont{Kiefl}},
  \bibinfo{author}{\bibfnamefont{R.~I.} \bibnamefont{Miller}},
  \bibnamefont{and} \bibinfo{author}{\bibfnamefont{G.~D.}
  \bibnamefont{Morris}}, \bibinfo{journal}{Science}
  \textbf{\bibinfo{volume}{292}}, \bibinfo{pages}{1692} (\bibinfo{year}{2001}).

\bibitem[{\citenamefont{Kaminski et~al.}(2002)\citenamefont{Kaminski,
  Rosenkranz, Fretwell, Campuzano, Li, Raffy, Cullen, You, Olson, Varma
  et~al.}}]{Kaminski:2002uj}
\bibinfo{author}{\bibfnamefont{A.}~\bibnamefont{Kaminski}},
  \bibinfo{author}{\bibfnamefont{S.}~\bibnamefont{Rosenkranz}},
  \bibinfo{author}{\bibfnamefont{H.~M.} \bibnamefont{Fretwell}},
  \bibinfo{author}{\bibfnamefont{J.~C.} \bibnamefont{Campuzano}},
  \bibinfo{author}{\bibfnamefont{Z.}~\bibnamefont{Li}},
  \bibinfo{author}{\bibfnamefont{H.}~\bibnamefont{Raffy}},
  \bibinfo{author}{\bibfnamefont{W.~G.} \bibnamefont{Cullen}},
  \bibinfo{author}{\bibfnamefont{H.}~\bibnamefont{You}},
  \bibinfo{author}{\bibfnamefont{C.~G.} \bibnamefont{Olson}},
  \bibinfo{author}{\bibfnamefont{C.~M.} \bibnamefont{Varma}},
  \bibnamefont{et~al.}, \bibinfo{journal}{Nature}
  \textbf{\bibinfo{volume}{416}}, \bibinfo{pages}{610} (\bibinfo{year}{2002}).

\bibitem[{\citenamefont{Borisenko et~al.}(2004)\citenamefont{Borisenko,
  Kordyuk, Koitzsch, Knupfer, Fink, Berger, and Lin}}]{Borisenko_et_al:2004}
\bibinfo{author}{\bibfnamefont{S.~V.} \bibnamefont{Borisenko}},
  \bibinfo{author}{\bibfnamefont{A.~A.} \bibnamefont{Kordyuk}},
  \bibinfo{author}{\bibfnamefont{A.}~\bibnamefont{Koitzsch}},
  \bibinfo{author}{\bibfnamefont{M.}~\bibnamefont{Knupfer}},
  \bibinfo{author}{\bibfnamefont{J.}~\bibnamefont{Fink}},
  \bibinfo{author}{\bibfnamefont{H.}~\bibnamefont{Berger}}, \bibnamefont{and}
  \bibinfo{author}{\bibfnamefont{C.~T.} \bibnamefont{Lin}},
  \bibinfo{journal}{Nature} \textbf{\bibinfo{volume}{431}}, \bibinfo{pages}{1}
  (\bibinfo{year}{2004}).

\bibitem[{\citenamefont{Arpiainen et~al.}(2009)\citenamefont{Arpiainen, Bansil,
  and Lindroos}}]{Arpiainen/Bansil/Lindroos:2009}
\bibinfo{author}{\bibfnamefont{V.}~\bibnamefont{Arpiainen}},
  \bibinfo{author}{\bibfnamefont{A.}~\bibnamefont{Bansil}}, \bibnamefont{and}
  \bibinfo{author}{\bibfnamefont{M.}~\bibnamefont{Lindroos}},
  \bibinfo{journal}{PRL} \textbf{\bibinfo{volume}{103}},
  \bibinfo{pages}{067005} (\bibinfo{year}{2009}).

\bibitem[{\citenamefont{Norman et~al.}(2010)\citenamefont{Norman, Kaminski, and
  Rosenkranz}}]{Norman/Kaminski/Rosenkranz:2010}
\bibinfo{author}{\bibfnamefont{M.~R.} \bibnamefont{Norman}},
  \bibinfo{author}{\bibfnamefont{A.}~\bibnamefont{Kaminski}}, \bibnamefont{and}
  \bibinfo{author}{\bibfnamefont{S.}~\bibnamefont{Rosenkranz}},
  \bibinfo{journal}{PRL} \textbf{\bibinfo{volume}{105}},
  \bibinfo{pages}{189701} (\bibinfo{year}{2010}).

\bibitem[{\citenamefont{Xia et~al.}(2008)\citenamefont{Xia, Schemm, Deutscher,
  Kivelson, Bonn, Hardy, Liang, Siemons, Koster, Fejer et~al.}}]{Xia:2008cc}
\bibinfo{author}{\bibfnamefont{J.}~\bibnamefont{Xia}},
  \bibinfo{author}{\bibfnamefont{E.}~\bibnamefont{Schemm}},
  \bibinfo{author}{\bibfnamefont{G.}~\bibnamefont{Deutscher}},
  \bibinfo{author}{\bibfnamefont{S.~A.} \bibnamefont{Kivelson}},
  \bibinfo{author}{\bibfnamefont{D.~A.} \bibnamefont{Bonn}},
  \bibinfo{author}{\bibfnamefont{W.~N.} \bibnamefont{Hardy}},
  \bibinfo{author}{\bibfnamefont{R.}~\bibnamefont{Liang}},
  \bibinfo{author}{\bibfnamefont{W.}~\bibnamefont{Siemons}},
  \bibinfo{author}{\bibfnamefont{G.}~\bibnamefont{Koster}},
  \bibinfo{author}{\bibfnamefont{M.~M.} \bibnamefont{Fejer}},
  \bibnamefont{et~al.}, \bibinfo{journal}{Phys. Rev. Lett.}
  \textbf{\bibinfo{volume}{100}}, \bibinfo{pages}{127002}
  (\bibinfo{year}{2008}).

\bibitem[{\citenamefont{MacDougall et~al.}(2008)\citenamefont{MacDougall,
  Aczel, Carlo, Ito, Rodriguez, Russo, Uemura, Wakimoto, and
  Luke}}]{MacDougall:2008br}
\bibinfo{author}{\bibfnamefont{G.~J.} \bibnamefont{MacDougall}},
  \bibinfo{author}{\bibfnamefont{A.~A.} \bibnamefont{Aczel}},
  \bibinfo{author}{\bibfnamefont{J.~P.} \bibnamefont{Carlo}},
  \bibinfo{author}{\bibfnamefont{T.}~\bibnamefont{Ito}},
  \bibinfo{author}{\bibfnamefont{J.}~\bibnamefont{Rodriguez}},
  \bibinfo{author}{\bibfnamefont{P.~L.} \bibnamefont{Russo}},
  \bibinfo{author}{\bibfnamefont{Y.~J.} \bibnamefont{Uemura}},
  \bibinfo{author}{\bibfnamefont{S.}~\bibnamefont{Wakimoto}}, \bibnamefont{and}
  \bibinfo{author}{\bibfnamefont{G.~M.} \bibnamefont{Luke}},
  \bibinfo{journal}{Phys. Rev. Lett.} \textbf{\bibinfo{volume}{101}},
  \bibinfo{pages}{017001} (\bibinfo{year}{2008}).

\bibitem[{\citenamefont{Sonier et~al.}(2009)\citenamefont{Sonier, Pacradouni,
  Sabok-Sayr, Hardy, Bonn, Liang, and Mook}}]{Sonier:2009fp}
\bibinfo{author}{\bibfnamefont{J.~E.} \bibnamefont{Sonier}},
  \bibinfo{author}{\bibfnamefont{V.}~\bibnamefont{Pacradouni}},
  \bibinfo{author}{\bibfnamefont{S.~A.} \bibnamefont{Sabok-Sayr}},
  \bibinfo{author}{\bibfnamefont{W.~N.} \bibnamefont{Hardy}},
  \bibinfo{author}{\bibfnamefont{D.~A.} \bibnamefont{Bonn}},
  \bibinfo{author}{\bibfnamefont{R.}~\bibnamefont{Liang}}, \bibnamefont{and}
  \bibinfo{author}{\bibfnamefont{H.~A.} \bibnamefont{Mook}},
  \bibinfo{journal}{Phys. Rev. Lett.} \textbf{\bibinfo{volume}{103}},
  \bibinfo{pages}{167002} (\bibinfo{year}{2009}).

\bibitem[{\citenamefont{Str{\"a}ssle et~al.}(2008)\citenamefont{Str{\"a}ssle,
  Roos, Mali, Keller, and Ohno}}]{Strassle:2008ku}
\bibinfo{author}{\bibfnamefont{S.}~\bibnamefont{Str{\"a}ssle}},
  \bibinfo{author}{\bibfnamefont{J.}~\bibnamefont{Roos}},
  \bibinfo{author}{\bibfnamefont{M.}~\bibnamefont{Mali}},
  \bibinfo{author}{\bibfnamefont{H.}~\bibnamefont{Keller}}, \bibnamefont{and}
  \bibinfo{author}{\bibfnamefont{T.}~\bibnamefont{Ohno}},
  \bibinfo{journal}{Phys. Rev. Lett.} \textbf{\bibinfo{volume}{101}},
  \bibinfo{pages}{237001} (\bibinfo{year}{2008}).

\bibitem[{\citenamefont{Varma}(1997)}]{Varma:1997}
\bibinfo{author}{\bibfnamefont{C.~M.} \bibnamefont{Varma}},
  \bibinfo{journal}{Phys. Rev. B} \textbf{\bibinfo{volume}{55}},
  \bibinfo{pages}{14554} (\bibinfo{year}{1997}).

\bibitem[{\citenamefont{Lovesey}(2014)}]{Lovesey:2014gp}
\bibinfo{author}{\bibfnamefont{S.~W.} \bibnamefont{Lovesey}},
  \bibinfo{journal}{J. Phys. Condens. Matter} \textbf{\bibinfo{volume}{26}},
  \bibinfo{pages}{356001} (\bibinfo{year}{2014}).

\bibitem[{\citenamefont{Lovesey et~al.}(2015)\citenamefont{Lovesey, Khalyavin,
  and Staub}}]{Lovesey:2014vm}
\bibinfo{author}{\bibfnamefont{S.~W.} \bibnamefont{Lovesey}},
  \bibinfo{author}{\bibfnamefont{D.~D.} \bibnamefont{Khalyavin}},
  \bibnamefont{and} \bibinfo{author}{\bibfnamefont{U.}~\bibnamefont{Staub}},
  \bibinfo{journal}{J. Phys. Condens. Matter} \textbf{\bibinfo{volume}{27}},
  \bibinfo{pages}{292201} (\bibinfo{year}{2015}).

\bibitem[{\citenamefont{Spaldin et~al.}(2013)\citenamefont{Spaldin, Fechner,
  Bousquet, Balatsky, and Nordstr\"om}}]{Spaldin_et_al:2013}
\bibinfo{author}{\bibfnamefont{N.~A.} \bibnamefont{Spaldin}},
  \bibinfo{author}{\bibfnamefont{M.}~\bibnamefont{Fechner}},
  \bibinfo{author}{\bibfnamefont{E.}~\bibnamefont{Bousquet}},
  \bibinfo{author}{\bibfnamefont{A.}~\bibnamefont{Balatsky}}, \bibnamefont{and}
  \bibinfo{author}{\bibfnamefont{L.}~\bibnamefont{Nordstr\"om}},
  \bibinfo{journal}{Physical Review B} \textbf{\bibinfo{volume}{88}},
  \bibinfo{pages}{094429} (\bibinfo{year}{2013}).

\bibitem[{\citenamefont{Mangin-Thro et~al.}(2015)\citenamefont{Mangin-Thro,
  Sidis, Wildes, and Bourges}}]{ManginThro_et_al:2015}
\bibinfo{author}{\bibfnamefont{L.}~\bibnamefont{Mangin-Thro}},
  \bibinfo{author}{\bibfnamefont{Y.}~\bibnamefont{Sidis}},
  \bibinfo{author}{\bibfnamefont{A.}~\bibnamefont{Wildes}}, \bibnamefont{and}
  \bibinfo{author}{\bibfnamefont{P.}~\bibnamefont{Bourges}},
  \bibinfo{journal}{Nature Commun.} \textbf{\bibinfo{volume}{6}},
  \bibinfo{pages}{7705} (\bibinfo{year}{2015}).

\bibitem[{\citenamefont{Orenstein}(2011)}]{Orenstein:2011}
\bibinfo{author}{\bibfnamefont{J.}~\bibnamefont{Orenstein}},
  \bibinfo{journal}{Phys. Rev. Lett.} \textbf{\bibinfo{volume}{107}},
  \bibinfo{pages}{067002} (\bibinfo{year}{2011}).

\bibitem[{\citenamefont{Barisic et~al.}(2008)\citenamefont{Barisic, Li, Zhao,
  Cho, Chabot-Couture, Yu, and Greven}}]{Barisic_et_al:2008}
\bibinfo{author}{\bibfnamefont{N.}~\bibnamefont{Barisic}},
  \bibinfo{author}{\bibfnamefont{Y.}~\bibnamefont{Li}},
  \bibinfo{author}{\bibfnamefont{X.}~\bibnamefont{Zhao}},
  \bibinfo{author}{\bibfnamefont{Y.-C.} \bibnamefont{Cho}},
  \bibinfo{author}{\bibfnamefont{G.}~\bibnamefont{Chabot-Couture}},
  \bibinfo{author}{\bibfnamefont{G.}~\bibnamefont{Yu}}, \bibnamefont{and}
  \bibinfo{author}{\bibfnamefont{M.}~\bibnamefont{Greven}},
  \bibinfo{journal}{Phys. Rev. B} \textbf{\bibinfo{volume}{78}},
  \bibinfo{pages}{054518} (\bibinfo{year}{2008}).

\bibitem[{\citenamefont{Yamamoto et~al.}(2000)\citenamefont{Yamamoto, Hu, and
  Tajima}}]{Yamamoto/Hu/Tajima:2000}
\bibinfo{author}{\bibfnamefont{A.}~\bibnamefont{Yamamoto}},
  \bibinfo{author}{\bibfnamefont{W.~Z.} \bibnamefont{Hu}}, \bibnamefont{and}
  \bibinfo{author}{\bibfnamefont{S.}~\bibnamefont{Tajima}},
  \bibinfo{journal}{Phys. Rev. B} \textbf{\bibinfo{volume}{63}},
  \bibinfo{pages}{024504} (\bibinfo{year}{2000}).

\bibitem[{\citenamefont{Huang et~al.}(1995)\citenamefont{Huang, Lynn, Xiong,
  and Chu}}]{Huang:1995tn}
\bibinfo{author}{\bibfnamefont{Q.}~\bibnamefont{Huang}},
  \bibinfo{author}{\bibfnamefont{J.~W.} \bibnamefont{Lynn}},
  \bibinfo{author}{\bibfnamefont{Q.}~\bibnamefont{Xiong}}, \bibnamefont{and}
  \bibinfo{author}{\bibfnamefont{C.~W.} \bibnamefont{Chu}},
  \bibinfo{journal}{Phys. Rev. B.} \textbf{\bibinfo{volume}{52}},
  \bibinfo{pages}{462} (\bibinfo{year}{1995}).

\bibitem[{\citenamefont{Itoh et~al.}(1996)\citenamefont{Itoh, Machi, Fukuoka,
  Tanabe, and Yasuoka}}]{Itoh_et_al:1996}
\bibinfo{author}{\bibfnamefont{Y.}~\bibnamefont{Itoh}},
  \bibinfo{author}{\bibfnamefont{T.}~\bibnamefont{Machi}},
  \bibinfo{author}{\bibfnamefont{A.}~\bibnamefont{Fukuoka}},
  \bibinfo{author}{\bibfnamefont{K.}~\bibnamefont{Tanabe}}, \bibnamefont{and}
  \bibinfo{author}{\bibfnamefont{H.}~\bibnamefont{Yasuoka}},
  \bibinfo{journal}{J. Phys. Soc. Jpn.} \textbf{\bibinfo{volume}{65}},
  \bibinfo{pages}{3751} (\bibinfo{year}{1996}).

\bibitem[{\citenamefont{Mounce et~al.}(2013)\citenamefont{Mounce, Oh, Lee,
  Halperin, Reyes, Kuhns, Chan, Dorow, Ji, Xia et~al.}}]{Mounce_et_al:2013}
\bibinfo{author}{\bibfnamefont{A.~M.} \bibnamefont{Mounce}},
  \bibinfo{author}{\bibfnamefont{S.}~\bibnamefont{Oh}},
  \bibinfo{author}{\bibfnamefont{J.~A.} \bibnamefont{Lee}},
  \bibinfo{author}{\bibfnamefont{W.~P.} \bibnamefont{Halperin}},
  \bibinfo{author}{\bibfnamefont{A.~P.} \bibnamefont{Reyes}},
  \bibinfo{author}{\bibfnamefont{P.~L.} \bibnamefont{Kuhns}},
  \bibinfo{author}{\bibfnamefont{M.~K.} \bibnamefont{Chan}},
  \bibinfo{author}{\bibfnamefont{C.}~\bibnamefont{Dorow}},
  \bibinfo{author}{\bibfnamefont{L.}~\bibnamefont{Ji}},
  \bibinfo{author}{\bibfnamefont{D.}~\bibnamefont{Xia}}, \bibnamefont{et~al.},
  \bibinfo{journal}{PRL} \textbf{\bibinfo{volume}{111}},
  \bibinfo{pages}{187003} (\bibinfo{year}{2013}).

\bibitem[{\citenamefont{Uchiyama et~al.}(2000)\citenamefont{Uchiyama, Hu,
  Yamamoto, Tajima, and Saiki}}]{Uchiyama_et_al:2000}
\bibinfo{author}{\bibfnamefont{H.}~\bibnamefont{Uchiyama}},
  \bibinfo{author}{\bibfnamefont{W.~Z.} \bibnamefont{Hu}},
  \bibinfo{author}{\bibfnamefont{A.}~\bibnamefont{Yamamoto}},
  \bibinfo{author}{\bibfnamefont{S.}~\bibnamefont{Tajima}}, \bibnamefont{and}
  \bibinfo{author}{\bibfnamefont{K.}~\bibnamefont{Saiki}},
  \bibinfo{journal}{Phys. Rev. B} \textbf{\bibinfo{volume}{62}},
  \bibinfo{pages}{615} (\bibinfo{year}{2000}).

\bibitem[{\citenamefont{Li et~al.}(2008)\citenamefont{Li, Bal{\'e}dent, Bari{\v
  s}i{\'c}, Cho, Fauqu{\'e}, Sidis, Yu, Zhao, Bourges, and
  Greven}}]{Li_et_al:2008}
\bibinfo{author}{\bibfnamefont{Y.}~\bibnamefont{Li}},
  \bibinfo{author}{\bibfnamefont{V.}~\bibnamefont{Bal{\'e}dent}},
  \bibinfo{author}{\bibfnamefont{N.}~\bibnamefont{Bari{\v s}i{\'c}}},
  \bibinfo{author}{\bibfnamefont{Y.}~\bibnamefont{Cho}},
  \bibinfo{author}{\bibfnamefont{B.}~\bibnamefont{Fauqu{\'e}}},
  \bibinfo{author}{\bibfnamefont{Y.}~\bibnamefont{Sidis}},
  \bibinfo{author}{\bibfnamefont{G.}~\bibnamefont{Yu}},
  \bibinfo{author}{\bibfnamefont{X.}~\bibnamefont{Zhao}},
  \bibinfo{author}{\bibfnamefont{P.}~\bibnamefont{Bourges}}, \bibnamefont{and}
  \bibinfo{author}{\bibfnamefont{M.}~\bibnamefont{Greven}},
  \bibinfo{journal}{Nature} \textbf{\bibinfo{volume}{455}},
  \bibinfo{pages}{372} (\bibinfo{year}{2008}).

\bibitem[{\citenamefont{Li et~al.}(2012)\citenamefont{Li, Yu, Chan,
  Bal{\'e}dent, Li, Bari{\v s}i{\'c}, Zhao, Hradil, Mole, Sidis
  et~al.}}]{Li:2012is}
\bibinfo{author}{\bibfnamefont{Y.}~\bibnamefont{Li}},
  \bibinfo{author}{\bibfnamefont{G.}~\bibnamefont{Yu}},
  \bibinfo{author}{\bibfnamefont{M.~K.} \bibnamefont{Chan}},
  \bibinfo{author}{\bibfnamefont{V.}~\bibnamefont{Bal{\'e}dent}},
  \bibinfo{author}{\bibfnamefont{Y.}~\bibnamefont{Li}},
  \bibinfo{author}{\bibfnamefont{N.}~\bibnamefont{Bari{\v s}i{\'c}}},
  \bibinfo{author}{\bibfnamefont{X.}~\bibnamefont{Zhao}},
  \bibinfo{author}{\bibfnamefont{K.}~\bibnamefont{Hradil}},
  \bibinfo{author}{\bibfnamefont{R.~A.} \bibnamefont{Mole}},
  \bibinfo{author}{\bibfnamefont{Y.}~\bibnamefont{Sidis}},
  \bibnamefont{et~al.}, \bibinfo{journal}{Nature Phys.}
  \textbf{\bibinfo{volume}{8}}, \bibinfo{pages}{404} (\bibinfo{year}{2012}).

\bibitem[{\citenamefont{Lovesey and Khalyavin}(2015)}]{Lovesey:2015wb}
\bibinfo{author}{\bibfnamefont{S.~W.} \bibnamefont{Lovesey}} \bibnamefont{and}
  \bibinfo{author}{\bibfnamefont{D.~D.} \bibnamefont{Khalyavin}},
  \bibinfo{journal}{arXiv} p. \bibinfo{pages}{1506.06859v1}
  (\bibinfo{year}{2015}).

\bibitem[{Bou()}]{BourgesPC}
\bibinfo{note}{P. Bourges, private communication}.

\bibitem[{\citenamefont{Yang et~al.}(2009)\citenamefont{Yang, Hwang,
  Schachinger, Carbotte, Lobo, Colson, Forget, and Timusk}}]{Yang_et_al:2009}
\bibinfo{author}{\bibfnamefont{J.}~\bibnamefont{Yang}},
  \bibinfo{author}{\bibfnamefont{J.}~\bibnamefont{Hwang}},
  \bibinfo{author}{\bibfnamefont{E.}~\bibnamefont{Schachinger}},
  \bibinfo{author}{\bibfnamefont{J.~P.} \bibnamefont{Carbotte}},
  \bibinfo{author}{\bibfnamefont{R.~P. S.~M.} \bibnamefont{Lobo}},
  \bibinfo{author}{\bibfnamefont{D.}~\bibnamefont{Colson}},
  \bibinfo{author}{\bibfnamefont{A.}~\bibnamefont{Forget}}, \bibnamefont{and}
  \bibinfo{author}{\bibfnamefont{T.}~\bibnamefont{Timusk}},
  \bibinfo{journal}{Phys. Rev. Lett.} \textbf{\bibinfo{volume}{102}},
  \bibinfo{pages}{027003} (\bibinfo{year}{2009}).

\bibitem[{\citenamefont{Tabis et~al.}(2014)\citenamefont{Tabis, Li, Le~Tacon,
  Braicovich, Kreyssig, Minola, Dellea, Weschke, Veit, Ramazanoglu
  et~al.}}]{Tabis:2014kb}
\bibinfo{author}{\bibfnamefont{W.}~\bibnamefont{Tabis}},
  \bibinfo{author}{\bibfnamefont{Y.}~\bibnamefont{Li}},
  \bibinfo{author}{\bibfnamefont{M.}~\bibnamefont{Le~Tacon}},
  \bibinfo{author}{\bibfnamefont{L.}~\bibnamefont{Braicovich}},
  \bibinfo{author}{\bibfnamefont{A.}~\bibnamefont{Kreyssig}},
  \bibinfo{author}{\bibfnamefont{M.}~\bibnamefont{Minola}},
  \bibinfo{author}{\bibfnamefont{G.}~\bibnamefont{Dellea}},
  \bibinfo{author}{\bibfnamefont{E.}~\bibnamefont{Weschke}},
  \bibinfo{author}{\bibfnamefont{M.~J.} \bibnamefont{Veit}},
  \bibinfo{author}{\bibfnamefont{M.}~\bibnamefont{Ramazanoglu}},
  \bibnamefont{et~al.}, \bibinfo{journal}{Nature Commun.}
  \textbf{\bibinfo{volume}{5}}, \bibinfo{pages}{1} (\bibinfo{year}{2014}).

\bibitem[{\citenamefont{Di~Matteo and Norman}(2012)}]{DiMatteo/Norman:2012}
\bibinfo{author}{\bibfnamefont{S.}~\bibnamefont{Di~Matteo}} \bibnamefont{and}
  \bibinfo{author}{\bibfnamefont{M.~R.} \bibnamefont{Norman}},
  \bibinfo{journal}{Phys. Rev. B.} \textbf{\bibinfo{volume}{85}},
  \bibinfo{pages}{235143} (\bibinfo{year}{2012}).

\bibitem[{\citenamefont{Yakovenko}(2015)}]{Yakovenko:2015}
\bibinfo{author}{\bibfnamefont{V.~M.} \bibnamefont{Yakovenko}},
  \bibinfo{journal}{Physica B: Phys. Condens. Matter}
  \textbf{\bibinfo{volume}{460}}, \bibinfo{pages}{159} (\bibinfo{year}{2015}).

\bibitem[{\citenamefont{Anisimov et~al.}(1991)\citenamefont{Anisimov, Zaanen,
  and Andersen}}]{Anisimov:1991wt}
\bibinfo{author}{\bibfnamefont{V.~I.} \bibnamefont{Anisimov}},
  \bibinfo{author}{\bibfnamefont{J.}~\bibnamefont{Zaanen}}, \bibnamefont{and}
  \bibinfo{author}{\bibfnamefont{O.~K.} \bibnamefont{Andersen}},
  \bibinfo{journal}{Phys. Rev. B.} \textbf{\bibinfo{volume}{44}},
  \bibinfo{pages}{943} (\bibinfo{year}{1991}).

\bibitem[{\citenamefont{Anisimov et~al.}(1997)\citenamefont{Anisimov,
  Aryasetiawan, and Liechtenstein}}]{Anisimov/Aryasetiawan/Liechtenstein:1997}
\bibinfo{author}{\bibfnamefont{V.~I.} \bibnamefont{Anisimov}},
  \bibinfo{author}{\bibfnamefont{F.}~\bibnamefont{Aryasetiawan}},
  \bibnamefont{and} \bibinfo{author}{\bibfnamefont{A.~I.}
  \bibnamefont{Liechtenstein}}, \bibinfo{journal}{J.~Phys.: Condens. Matter}
  \textbf{\bibinfo{volume}{9}}, \bibinfo{pages}{767} (\bibinfo{year}{1997}).

\bibitem[{\citenamefont{Wu et~al.}(2006)\citenamefont{Wu, Zhang, and
  Tao}}]{Wu:2006gj}
\bibinfo{author}{\bibfnamefont{D.}~\bibnamefont{Wu}},
  \bibinfo{author}{\bibfnamefont{Q.}~\bibnamefont{Zhang}}, \bibnamefont{and}
  \bibinfo{author}{\bibfnamefont{M.}~\bibnamefont{Tao}},
  \bibinfo{journal}{Phys. Rev. B.} \textbf{\bibinfo{volume}{73}},
  \bibinfo{pages}{235206} (\bibinfo{year}{2006}).

\bibitem[{\citenamefont{Himmetoglu et~al.}(2011)\citenamefont{Himmetoglu,
  Wentzcovitch, and Cococcioni}}]{Himmetoglu:2011gn}
\bibinfo{author}{\bibfnamefont{B.}~\bibnamefont{Himmetoglu}},
  \bibinfo{author}{\bibfnamefont{R.~M.} \bibnamefont{Wentzcovitch}},
  \bibnamefont{and}
  \bibinfo{author}{\bibfnamefont{M.}~\bibnamefont{Cococcioni}},
  \bibinfo{journal}{Phys. Rev. B.} \textbf{\bibinfo{volume}{84}},
  \bibinfo{pages}{115108} (\bibinfo{year}{2011}).

\bibitem[{\citenamefont{Kresse and Furthm{\"u}ller}(1996)}]{Kresse:1996vf}
\bibinfo{author}{\bibfnamefont{G.}~\bibnamefont{Kresse}} \bibnamefont{and}
  \bibinfo{author}{\bibfnamefont{J.}~\bibnamefont{Furthm{\"u}ller}},
  \bibinfo{journal}{Phys. Rev. B.} \textbf{\bibinfo{volume}{54}},
  \bibinfo{pages}{11169} (\bibinfo{year}{1996}).

\bibitem[{\citenamefont{Bl{\"o}chl}(1994)}]{Blochl:1994uk}
\bibinfo{author}{\bibfnamefont{P.}~\bibnamefont{Bl{\"o}chl}},
  \bibinfo{journal}{Phys. Rev. B.} \textbf{\bibinfo{volume}{50}},
  \bibinfo{pages}{17953} (\bibinfo{year}{1994}).

\bibitem[{ELK()}]{ELK}
\emph{\bibinfo{title}{\uppercase{\textnormal{lapw}}
  \uppercase{\textnormal{elk}} code}},
  \bibinfo{note}{http://elk.sourceforge.net}.

\bibitem[{\citenamefont{Ambrosch-Draxl
  et~al.}(2003)\citenamefont{Ambrosch-Draxl, S{\"u}le, Auer, and
  Sherman}}]{AmbroschDraxl:2003ff}
\bibinfo{author}{\bibfnamefont{C.}~\bibnamefont{Ambrosch-Draxl}},
  \bibinfo{author}{\bibfnamefont{P.}~\bibnamefont{S{\"u}le}},
  \bibinfo{author}{\bibfnamefont{H.}~\bibnamefont{Auer}}, \bibnamefont{and}
  \bibinfo{author}{\bibfnamefont{E.}~\bibnamefont{Sherman}},
  \bibinfo{journal}{Phys. Rev. B.} \textbf{\bibinfo{volume}{67}},
  \bibinfo{pages}{100505} (\bibinfo{year}{2003}).

\bibitem[{\citenamefont{Thonhauser et~al.}(2004)\citenamefont{Thonhauser, Auer,
  Sherman, and Ambrosch-Draxl}}]{Thonhauser:2004jr}
\bibinfo{author}{\bibfnamefont{T.}~\bibnamefont{Thonhauser}},
  \bibinfo{author}{\bibfnamefont{H.}~\bibnamefont{Auer}},
  \bibinfo{author}{\bibfnamefont{E.}~\bibnamefont{Sherman}}, \bibnamefont{and}
  \bibinfo{author}{\bibfnamefont{C.}~\bibnamefont{Ambrosch-Draxl}},
  \bibinfo{journal}{Phys. Rev. B.} \textbf{\bibinfo{volume}{69}},
  \bibinfo{pages}{104508} (\bibinfo{year}{2004}).

\bibitem[{\citenamefont{Ambrosch-Draxl
  et~al.}(2004{\natexlab{a}})\citenamefont{Ambrosch-Draxl, Sherman, Auer, and
  Thonhauser}}]{AmbroschDraxl:2004kp}
\bibinfo{author}{\bibfnamefont{C.}~\bibnamefont{Ambrosch-Draxl}},
  \bibinfo{author}{\bibfnamefont{E.~Y.} \bibnamefont{Sherman}},
  \bibinfo{author}{\bibfnamefont{H.}~\bibnamefont{Auer}}, \bibnamefont{and}
  \bibinfo{author}{\bibfnamefont{T.}~\bibnamefont{Thonhauser}},
  \bibinfo{journal}{phys. stat. sol. (b)} \textbf{\bibinfo{volume}{241}},
  \bibinfo{pages}{1199} (\bibinfo{year}{2004}{\natexlab{a}}).

\bibitem[{\citenamefont{Ambrosch-Draxl
  et~al.}(2004{\natexlab{b}})\citenamefont{Ambrosch-Draxl, Sherman, Auer, and
  Thonhauser}}]{AmbroschDraxl:2004bt}
\bibinfo{author}{\bibfnamefont{C.}~\bibnamefont{Ambrosch-Draxl}},
  \bibinfo{author}{\bibfnamefont{E.}~\bibnamefont{Sherman}},
  \bibinfo{author}{\bibfnamefont{H.}~\bibnamefont{Auer}}, \bibnamefont{and}
  \bibinfo{author}{\bibfnamefont{T.}~\bibnamefont{Thonhauser}},
  \bibinfo{journal}{Phys. Rev. Lett.} \textbf{\bibinfo{volume}{92}},
  \bibinfo{pages}{187004} (\bibinfo{year}{2004}{\natexlab{b}}).

\bibitem[{\citenamefont{Ambrosch-Draxl and
  Sherman}(2006)}]{AmbroschDraxl:2006jc}
\bibinfo{author}{\bibfnamefont{C.}~\bibnamefont{Ambrosch-Draxl}}
  \bibnamefont{and} \bibinfo{author}{\bibfnamefont{E.}~\bibnamefont{Sherman}},
  \bibinfo{journal}{Phys. Rev. B.} \textbf{\bibinfo{volume}{74}},
  \bibinfo{pages}{024503} (\bibinfo{year}{2006}).

\bibitem[{\citenamefont{Munoz et~al.}(2000)\citenamefont{Munoz, Illas, and
  Moreira}}]{Anonymous:ZSMerAaE}
\bibinfo{author}{\bibfnamefont{D.}~\bibnamefont{Munoz}},
  \bibinfo{author}{\bibfnamefont{F.}~\bibnamefont{Illas}}, \bibnamefont{and}
  \bibinfo{author}{\bibfnamefont{I.}~\bibnamefont{Moreira}},
  \bibinfo{journal}{Phys. Rev. Lett.} \textbf{\bibinfo{volume}{84}},
  \bibinfo{pages}{1579} (\bibinfo{year}{2000}).

\bibitem[{\citenamefont{d'Astuto et~al.}(2003)\citenamefont{d'Astuto, Mirone,
  Giura, Colson, Forget, and Krisch}}]{dAstuto:2003cy}
\bibinfo{author}{\bibfnamefont{M.}~\bibnamefont{d'Astuto}},
  \bibinfo{author}{\bibfnamefont{A.}~\bibnamefont{Mirone}},
  \bibinfo{author}{\bibfnamefont{P.}~\bibnamefont{Giura}},
  \bibinfo{author}{\bibfnamefont{D.}~\bibnamefont{Colson}},
  \bibinfo{author}{\bibfnamefont{A.}~\bibnamefont{Forget}}, \bibnamefont{and}
  \bibinfo{author}{\bibfnamefont{M.}~\bibnamefont{Krisch}},
  \bibinfo{journal}{J. Phys. Condens. Matter} \textbf{\bibinfo{volume}{15}},
  \bibinfo{pages}{8827} (\bibinfo{year}{2003}).

\bibitem[{\citenamefont{Uchiyama et~al.}(2004)\citenamefont{Uchiyama, Baron,
  Tsutsui, Tanaka, Hu, Yamamoto, Tajima, and Endoh}}]{Uchiyama:2004kl}
\bibinfo{author}{\bibfnamefont{H.}~\bibnamefont{Uchiyama}},
  \bibinfo{author}{\bibfnamefont{A.}~\bibnamefont{Baron}},
  \bibinfo{author}{\bibfnamefont{S.}~\bibnamefont{Tsutsui}},
  \bibinfo{author}{\bibfnamefont{Y.}~\bibnamefont{Tanaka}},
  \bibinfo{author}{\bibfnamefont{W.~Z.} \bibnamefont{Hu}},
  \bibinfo{author}{\bibfnamefont{A.}~\bibnamefont{Yamamoto}},
  \bibinfo{author}{\bibfnamefont{S.}~\bibnamefont{Tajima}}, \bibnamefont{and}
  \bibinfo{author}{\bibfnamefont{Y.}~\bibnamefont{Endoh}},
  \bibinfo{journal}{Phys. Rev. Lett.} \textbf{\bibinfo{volume}{92}},
  \bibinfo{pages}{197005} (\bibinfo{year}{2004}).

\bibitem[{num()}]{numerical}
\bibinfo{note}{We use the numerical solver for differential equations
  implemented in the Mathematica \cite{mathematica} software package. For the
  friction constant $\gamma$ we use the lifetime of the phonon
  ($\tau=1/\gamma$), which we set to a value of \unit[0.25]{ps}. The numerical
  solution requires two initial conditions, which we take to be a non-displaced
  mode ($Q(0)=0$) with a finite velocity ($Q'(0)\neq0$).}

\bibitem[{\citenamefont{Krichevtsov et~al.}(1993)\citenamefont{Krichevtsov,
  Pavlov, Pisarev, and Gridnev}}]{Krichevtsov:1993gs}
\bibinfo{author}{\bibfnamefont{B.~B.} \bibnamefont{Krichevtsov}},
  \bibinfo{author}{\bibfnamefont{V.~V.} \bibnamefont{Pavlov}},
  \bibinfo{author}{\bibfnamefont{R.~V.} \bibnamefont{Pisarev}},
  \bibnamefont{and} \bibinfo{author}{\bibfnamefont{V.~N.}
  \bibnamefont{Gridnev}}, \bibinfo{journal}{J. Phys. Condens. Matter}
  \textbf{\bibinfo{volume}{5}}, \bibinfo{pages}{8233} (\bibinfo{year}{1993}).

\bibitem[{\citenamefont{Timusk and Statt}(1999)}]{Timusk:1999wp}
\bibinfo{author}{\bibfnamefont{T.}~\bibnamefont{Timusk}} \bibnamefont{and}
  \bibinfo{author}{\bibfnamefont{B.}~\bibnamefont{Statt}},
  \bibinfo{journal}{Rep. Prog. Phys.} \textbf{\bibinfo{volume}{62}},
  \bibinfo{pages}{61} (\bibinfo{year}{1999}).

\bibitem[{\citenamefont{Hu et~al.}(2014)\citenamefont{Hu, Kaiser, Nicoletti,
  Hunt, Gierz, Hoffmann, Le~Tacon, Loew, Keimer, and
  Cavalleri}}]{Hu_et_al:2014}
\bibinfo{author}{\bibfnamefont{W.}~\bibnamefont{Hu}},
  \bibinfo{author}{\bibfnamefont{S.}~\bibnamefont{Kaiser}},
  \bibinfo{author}{\bibfnamefont{D.}~\bibnamefont{Nicoletti}},
  \bibinfo{author}{\bibfnamefont{C.~R.} \bibnamefont{Hunt}},
  \bibinfo{author}{\bibfnamefont{I.}~\bibnamefont{Gierz}},
  \bibinfo{author}{\bibfnamefont{M.~C.} \bibnamefont{Hoffmann}},
  \bibinfo{author}{\bibfnamefont{M.}~\bibnamefont{Le~Tacon}},
  \bibinfo{author}{\bibfnamefont{T.}~\bibnamefont{Loew}},
  \bibinfo{author}{\bibfnamefont{B.}~\bibnamefont{Keimer}}, \bibnamefont{and}
  \bibinfo{author}{\bibfnamefont{A.}~\bibnamefont{Cavalleri}},
  \bibinfo{journal}{Nat. Mater.} \textbf{\bibinfo{volume}{13}},
  \bibinfo{pages}{705} (\bibinfo{year}{2014}).

\bibitem[{\citenamefont{Mankowsky et~al.}(2014)\citenamefont{Mankowsky, Subedi,
  F\"orst, Mariager, Chollet, Lemke, Robinson, Glownia, Minitti, Frano
  et~al.}}]{Mankowsky_et_al:2014}
\bibinfo{author}{\bibfnamefont{R.}~\bibnamefont{Mankowsky}},
  \bibinfo{author}{\bibfnamefont{A.}~\bibnamefont{Subedi}},
  \bibinfo{author}{\bibfnamefont{M.}~\bibnamefont{F\"orst}},
  \bibinfo{author}{\bibfnamefont{S.~O.} \bibnamefont{Mariager}},
  \bibinfo{author}{\bibfnamefont{M.}~\bibnamefont{Chollet}},
  \bibinfo{author}{\bibfnamefont{H.~T.} \bibnamefont{Lemke}},
  \bibinfo{author}{\bibfnamefont{J.~S.} \bibnamefont{Robinson}},
  \bibinfo{author}{\bibfnamefont{J.~M.} \bibnamefont{Glownia}},
  \bibinfo{author}{\bibfnamefont{M.~P.} \bibnamefont{Minitti}},
  \bibinfo{author}{\bibfnamefont{A.}~\bibnamefont{Frano}},
  \bibnamefont{et~al.}, \bibinfo{journal}{Nature}
  \textbf{\bibinfo{volume}{516}}, \bibinfo{pages}{71} (\bibinfo{year}{2014}).

\end{thebibliography}

\end{document}